# Modelling the data and not the images in FMRI

**Thomas W. D. Möbius**[a]

[a]Institute of Medical Informatics and Statistics, Kiel University, Kiel, Germany

## Abstract

The standard approach to the analysis of functional magnetic resonance imaging (FMRI) data applies various preprocessing steps to the original FMRI. These preprocessings lead to a general underestimation of residual variance in the downstream analysis. This negatively impacts the type I error of statistical tests and increases the risk for reporting false positive results. A genuine approach to the statistical analysis of FMRI data of brain scans is derived from first principles that is deeply rooted in statistical test theory. The method combines all preprocessing steps of the standard approach into one single modelling step, enabling valid statistical tests to be constructed. On population level, BOLD effects are modelled by random effects meta regression models. This acknowledges that subjects are random entities, and it acknowledges that the accuracy of the BOLD signal is estimated with various certainty in an FMRI. The concept of a reference scalar field is introduced that enables individual effect sizes to be related to each other with respect to a common unit. In particular, multicentre studies will gain interpretability and power by its use.

## 1 Introduction

In their review of the implementation and use of the general linear model (GLM) in functional magnetic resonance imaging (FMRI) analysis, Poline and Brett (1) make the observation that the FMRI community relatively sparsely interacts with estimation theory experts, mathematicians, and model builders from the statistics community. This is not the first time this observation has been made. Already ten years ago, Lindquist (2) asked statisticians to study the estimation and decision-making process in FMRI from a specific statistical viewpoint. This invitation has just recently been renewed by Brown and Behrmann (3) after Eklund et al. (4) attested the current, state-of-the-art approaches to FMRI a severe inflation in false positives.

Beside its intriguing complexity, though, mathematical statisticians may be reluctant to join the field: The field gives the rather tight impression that the big picture of how and which substantial steps have to be taken in a FMRI data analysis have long been solved. The cascade of *motion and slice timing correction*, *normalisation*, *spatial smoothing*, and postpositioned *statistical modelling* of voxel-based time series data is approaching its 25th anniversary. The procedure appears to be set in stone as generations of researchers in the field have since then grown up with it. As the method is closely centred around the connotation that the statistical analysis is at its essence an analysis of sequentially gathered pictures, we shall refer to the method as the *picture based* (PB) approach to FMRI. If you are using SPM (5,6), FSL (7), or AFNI (8) to analyse an FMRI, then you are using the PB approach to estimate the respective BOLD effect in the data.

It appears that for statisticians and data scientists there are only details left to fill: improvements are made in image registration, spatial smoothing is moving from a more holistic whole-brain approach to more adaptive, iterated procedures (9), and there is, of course, a lively discussion on how to build, select, and validate models in FMRI. Beside these advances, the basic principle of the PB approach has remained unchanged.

The separation of the analysis into *data correction*, *normalisation*, *spatial smoothing*, and *statistical modelling* has one, arguably big disadvantage, though: variance is lost in the preprocessing. Any projection of data onto a regression surface smooths data and for statistical tests it is important to know the variance of the data around this surface: the residual variance. Knowledge of residual variance is with no exaggeration the key ingredient for the construction of valid statistical test. But in PB estimation, the statistical model only sees the temporal dimension of the spatio-temporal variation of the residuals: the spatial dimension is lost in the preprocessing.

This manuscript presents a small but subtle trick that combines *all* of the analysis steps of the PB estimator into *one* single model fitting step. (As the newly derived method is intrinsically model based, we shall refer to it as the *model based* (MB) approach to FMRI.) The method is derived from first principles, the estimator is, more than any current approaches, deeply rooted in statistical test theory, and the framework yields a clean, statistical methodology to FMRI data analysis. The approach enables access to the full spatio-temporal variance of the data in all data analysis steps, in model building, model validation, model selection, and last but not least, testing.

## 2 A general model for FMRI data

Given data and a statistical model for the data, an estimator finds appropriate values for the parameters of the model that best fit the observed data within the constraints of the model. Thus any statistical analysis involves at least these two choices: we have to choose a model for the data, and we have to choose an estimator that will create a *fit* of the model. For users, the difference between *modelling data by a model* and *estimating the parameters of a model* may appear fuzzy as both are tightly linked and always go hand-in-hand in all statistical analyses. If a statistical software package additionally hides the fitting process itself from the user, the difference may even further be blurred.

Statistical theory offers a variety of different utilities that allow to characterise estimators and to study their particular advantages and disadvantages. Among the most important are the concepts of *validity* and *power* of test statistics which have been constructed from an estimator (a test statistic is valid if it adheres to its type I error). Ex in statistical terms, Carp (10,11) showed that test statistics based on the PB estimator have inflated type I errors, and Eklund et al. (4) showed that this is because the variance of the PB estimator is underestimated.

This is a manuscript about estimation. Of course, without a model there is no need for an estimator, and albeit the intention is not to add new FMRI models to the literature, the discussion shall nevertheless start by deriving a modelling framework for the remainder of the text.

There are many experimental designs in FMRI. Among these, block designs have a relatively simple structure. In the following, a model for block design studies is defined as complex enough to allow to explain the features and aspects of the new estimator and to compare it to the standard PB approach. In block design studies, it is of interest to infer task related neural activity in the brain. Such activations may relate to specific motor functions, such as finger tapping, or may relate to characteristic processings of a brain when asked to perform certain mental tasks. During the experiment, brain images are acquired when the subject is presented with blocks of stimuli, say blocks of type $A$, and control blocks of type $B$. The question is whether there exist areas in which these images significantly change between these blocks.

Since an increase in blood oxygenation and flow into areas of the brain is a surrogate for increased brain activity (2), called the hemodynamic response, these areas can be interpreted as changes in neural activity when MR settings are chosen which are sensitive to changes in blood oxygenation, i.e. which target the blood oxygen level dependent (BOLD) contrast.

The aim of FMRI studies is thus twofold: (i) given series of MR signal images of a subject, the expected BOLD signal (the *activation field*) of the subject in this FMRI is to be inferred, and (ii) given a representative sample of activation fields from a population, the expected activation field of a randomly sampled individual from the population is to be inferred.

Although brains in a population are different, it is common to assume that they share enough characteristics that for each pair of brains there exists a diffeomorphic map between them. Let $M$ be a representative of this population of brains. Standard templates, say from the Montreal Neurological Institute, are typically chosen for $M$. (If interest only lies in estimating (i), $M$ may be set equal to the subject brain itself.) Mathematically speaking, it is assumed that there exists diffeomorphisms $\psi_j$ for all subjects $j$ in the population for which $\psi_j[M]$ is the brain of $j$.

Let $t$ denote time and $\mathbb{1}_t$ the (partially defined) indicator

$$\mathbb{1}_t = \begin{cases} 1 & \text{for } t \text{ within a block of type } A, \\ 0 & \text{for } t \text{ within a block of type } B, \\ & \text{undefined otherwise.} \end{cases}$$

Let $x \in M$ be a point in standard space, and let $\gamma(x)$ denote the expected task related signal change at this point, i.e. the mean signal difference between blocks one would expect if we were to sample a time series of MR signals at $\psi_j(x)$ in the brain of a randomly selected individual $j$ from the population. In contrast, given a fixed subject $j$, let $\beta_j(x)$ be the expected task related signal change at $\psi_j(x)$ in *this* subject and *this* FMRI, i.e. the expected block difference *conditional* that we already know that it is $j$ who lies in the scanner.

Let $Y_{jt}(x)$ denote the MR signal at $\psi_j(x)$ during $t$. Then $Y_{jt}(x)$, essentially, follows the model

$$Y_{jt}(x) = \alpha_j(x) + \mathbb{1}_t \cdot \beta_j(x) + f_j(t,x) + \varepsilon_{jt}(x), \qquad [1]$$

$$\beta_j(x) = \gamma(x) + z_j^\top \delta(x) + \xi_j(x) \qquad [2]$$

where $\varepsilon_{jt}(x)$ and $\xi_j(x), x \in M$ denote random scalar fields.

It is common to add further covariates to [1] or to model the covariate $\mathbb{1}_t$ as a convolution of delta functions with a canonical hemodynamic response function (1, 2). The random fields $\varepsilon_{jt}$ and $\xi_j$ model »unexplainable variances« in the data, i.e. the signal variations which are not captured by the deterministic parts of [1] and [2]. The residual field $\varepsilon_{jt}$ models the variability of the MR signal within the FMRI session of subject $j$. The residual field $\xi_j$ models the deviation of the $j$th subject activation field, namely $\beta_j(x)$, from the expected population average.

The functions $f_j(\cdot, x)$ model temporal fluctuations of the MR signal during the course of the experiment which are independent from the BOLD signal. If $f_j(\cdot, t)$ is specified correctly, the unexplained variance in the data is reduced and the power of tests on, e.g., $\beta_j(x)$ increases. Overfitting of $f_j(\cdot, x)$, however, will reduce this power.

Call $\alpha_j = (\alpha_j(x) : x \in M)$ the intercept field of the $j$th subject and $\beta_j = (\beta_j(x) : x \in M)$ the subject's activation field. To test whether there is a positive activation at $\psi_j(x)$ in the brain of subject $j$, we test

$$H_0(x) : \beta_j(x) = 0 \quad \text{vs.} \quad H_1(x) : \beta_j(x) > 0.$$

Potential covariates are modelled by the covariate vectors $z_j$, which may code for age, sex, or case/control status. The expected activation field of a randomly selected subject $j$ from the population is

$$\mathbb{E}\left(\beta_j(x)\right) = \gamma(x) + z_j^\top \delta(x).$$

If the population stratifies with respect to a single attribute, say case/control status, the stratification may be coded as a single dichotomous variable, say $z_j = 0$ for control and $z_j = 1$ for cases. Then testing for areas which differ between the strata corresponds to testing

$$H_0(x) : \delta(x) = 0 \quad \text{vs.} \quad H_1(x) : \delta(x) \neq 0.$$

If no covariates are of interest, the term $z_j^\top \delta$ can be dropped and testing for positive activations simplifies to

$$H_0(x) : \gamma(x) = 0 \quad \text{vs.} \quad H_1(x) : \gamma(x) > 0.$$

The difficulty in fitting the model [1]+[2] to data lies in the inability to repeatedly sample MR signals at any given, fixed set of points $x$ in $M$: During the course of an FMRI session, the MR scanner acquires series of 2D images, in which each image corresponds to one slice of a predefined acquisition grid. All slices of the grid are measured in fixed order until measurements start anew with the first slice in the succession. Refer to a set of successive scans which cover the acquisition grid as a *scan cycle*. All intensities of one slice are the result of the same measurement, and intensities of adjacent slices are shifted in time. In particular, measurements in the 3D image of a scan cycle are not measured simultaneously. Since a subject moves ever so slightly in the scanner, the location of a point in the acquisition grid, i.e. a point with fixed coordinates with respect to the scanner, will lie at a different location in the brain each time the respective slice is measured anew (Figure 1). Mathematically, we deal with families of rigid body transformations $\left(\rho_{jt} : t \in T\right)$ for each subject $j$ which map the brain $\psi_j[M]$ of $j$ into scanner space at time $t \in T$. As the $\rho_{jt}$ change continuously in $t$, no grid $\rho_{jt}^{-1}[V_{jt}]$ will coincide with any other grid at any other time during a session. In consequence, no point $x \in M$ is measured more than once in a study.

## 3 Fitting the signal model

From an imaging viewpoint, the analysis has to deal with a moving object behind a fixed lattice (Figure 1). The problem is consequently solved by first specifying a reference lattice with respect to the moving object, and then to spatially and temporally *realign* the original images onto this new grid. To realign or *to register* images onto a new grid means to interpolate[1] the original imaging data to all points of the new grid, and then to discard all data of the original images (Figure 2). This inevitably reduces variance in the data. As touched upon briefly in the introduction, this reduction in variance challenges the validity of downstream statistical tests. We shall come back to this issue.

Let us imagine, now, that it is in truth the subject's head which remains still over the course of the respective FMRI session, and let us instead picture the MR scanner as if it would move around the subject's head (Figure 1). Then seen from a statistical viewpoint, measurements are simply taken at different points in space and time.

It is physically justifiable to assume that the MR signal field is continuous. The closer a signal is measured to a point $x$ of interest, the closer the acquired signal should be to the *true* signal at $x$. This is not a new assumption. If the MR signal field were not continuous, interpolations as they are applied repeatedly by the PB estimator would be invalid. Consequently, when estimating the signal at a point $x$, measurements in close proximity to $x$ should gain more weight than measurements taken further apart. This suggests to estimate the (unobserved) MR signal at $x$ via a weighted average of the surrounding observations using a weighting scheme that reflects this increased trust. Let

$$\omega^x \colon \mathbb{R}^3 \to \mathbb{R}, \quad r \mapsto \omega_r^x$$

be such a weighting scheme for $x$. If $r$ is a point in the brain of a subject at which an actual MR signal has been measured, this measurement will get the weight $\omega_r^x$ when estimating the signal intensity at $x$. Measurements in the proximity of $x$ are $r = \rho_{jt}^{-1}(v)$ where $v$ is a voxel in the acquisition grid of the scanner and $\rho_{jt}$ is the rigid body transformation which maps the brain of subject $j$ to its current location at time $t$ (Figure 2). In [1], we denoted the MR signal at $\rho_{jt}^{-1}(v)$ during $t$ with

$$Y_{jt}\left(\rho_{jt}^{-1}(v)\right).$$

Let us denote the actual MR signal which has been measured at $\rho_{jt}^{-1}(v)$ during $t$ with

$$y_{jt}\left(\rho_{jt}^{-1}(v)\right).$$

And when estimating the signal at $x \in M$, this measurement gets the weight

$$\omega^{\psi_j(x)}_{\rho_{jt}^{-1}(v)}.$$

Hence, this naturally suggests to estimate the activation field $\beta_j(x), x \in M$ of the $j$th subject at a point $x \in M$ by solving the weighted least squares (WLS) optimisation problem

[1] In FMRI, interpolation is often referred to as *resampling*. In statistics, though, *resampling* is a fixed term exclusively used in the context of bootstrapping or Markov chain Monte Carlo (MCMC) samplings when samples are drawn from an empirical distribution by, say, a Metropolis-Hastings algorithm. As resampling has this very reserved meaning in statistics, I will use the term interpolation for interpolations instead.

$$e_{jt}(v,\alpha,\beta,f) = y_{jt}\left(\rho_t^{-1}(v)\right) - \alpha + \mathbb{1}_t \cdot \beta - f(t,v),$$

$$\hat{\mu}_j(x) = \begin{pmatrix} \hat{\alpha}_j(x) \\ \hat{\beta}_j(x) \\ \hat{f}_j(\cdot,x) \end{pmatrix} = \underset{\alpha,\beta,f}{\text{argmin}} \frac{\sum_{t,v} \omega^{\psi_j(x)}_{\rho_{jt}^{-1}(v)} e^2_{jt}(v,\alpha,\beta,f)}{\sum_{t,v} \omega^{\psi_j(x)}_{\rho_{jt}^{-1}(v)}}. \quad [3]$$

The sum $\sum_{t,v}$ in the formula runs over all $t$ at which $\mathbb{1}_t$ is defined and over all grid points $v \in V_{jt}$ in the respective slice that are measured during $t$.

The estimator results in a spatio-temporal smooth of the FMRI data onto a regression (hyper-)surface, here denoted by

$$\hat{\mu}_j(x), x \in M.$$

The $\hat{\beta}_j$ dimension of $\hat{\mu}_j$ is an estimate of the BOLD signal at $x$. The estimator combines the steps *normalisation*, *spatial smooth* and *voxel-centric time-series modelling* of the PB estimator into one single fit of a statistical model while making all data correction steps of the PB approach meaningless. There is no need to »correct« data for motion or slice timing differences as the respective information is directly *used* by the estimator. The estimator does not assume that the slices of a scan cycle have been measured simultaneously nor that any slice through the subject brain coincides with any other at any other time during the session. As [3] very directly estimates the parameters of the underlying FMRI model (here [1]), the method is referred to as the *model based* (MB) approach to FMRI.

The choice of the weighting scheme in [3] runs parallel to the choice of which kernel to use for the spatial smooth in PB estimation. Any advantages and disadvantages of using one weighting scheme over the other or the potential advantages of using iterated procedures for finding adequate weighting schemes equally apply to PB and MB estimation alike. The essential difference by which the MB estimator sets itself apart from the PB approach is its ability to estimate the actual spatio-temporal residual variance of the data at each $x$. This has direct consequences for the estimated variance of $\hat{\beta}_j(x)$,

$$\hat{\sigma}^2\left(\hat{\beta}_j(x)\right). \quad [4]$$

In PB estimation, the variance of $\hat{\beta}_j(x)$ is only estimated along the temporal dimension of the data – remember that spatial smooths are part of the PB preprocessing and the spatial dimension is decoupled from the actual voxel-based GLM.[2] A potentially large part of the variance of $\hat{\beta}_j(x)$ is thus hidden from the statistical model. Indeed, Figure 7 (a figure that will be introduced and discussed in detail) prompts that the spatial dimension of the residual variation is indeed quite essential for yielding proper estimates for [4]. The underestimation of $\sigma^2\left(\hat{\beta}_j(x)\right)$ in PB estimation may largely explain the observed type I error inflation of the approach.

The key selling point of the MB is the method's ability to provide actual estimates for [4] by construction. In MB estimation, t-test statistics fields

$$t_j(x) = \frac{\hat{\beta}_j(x)}{\hat{\sigma}\left(\hat{\beta}_j(x)\right)} \quad [5]$$

are provided with real, trustworthy estimates in the denominators at all points $x \in M$. This is a massive win for the validity of statistical tests.

Another difference to the standard approach is that activation fields and t-test statistics fields can be evaluated at any desirable point $x \in M$ in the template: the method is not voxel-centric. Evaluations may even be chosen on the fly. As points do not have to be chosen in advance, the MB estimator can be coupled with search algorithms which scan parameter fields for local optima or trace the hills and valleys of the these fields. This is impossible with the current voxel-centric approaches.

## 4 Random field theory

Basic requisites for weighting schemes $\omega^x_r$ in [3] are that weights should be positive, strictly monotone decreasing in $r - t$ (i.e. $\omega^x_r$ should decrease if the distance between $x$ and $r$ increases), and weights should vanish at infinity (i.e. if the distance of $x$ and $r$ is large, $\omega^x_r$ should approach or be identical to 0). Also, properties of continuity and differentiability are desirable: if the weighting schemes are jointly smooth in $x$ and $r$, then so are all realisations of $\hat{\beta}_j$. This suggests to call a weighting scheme $\omega = (\omega^x)_{x \in M}$ *valid* if for all $x \in M$: (i) $\omega^x \geq 0$, (ii) $\omega^x_r < \omega^x_s$ when $\|r - x\| > \|s - x\|$, (iii) $\omega^x_r \to 0$ for $\|r\| \to \infty$, and (iv) $\omega^x(r)$ is jointly smooth in $(x, r)$.

If $w$ is valid, the new MB estimator provides smooth fits of the respective parameters fields of [1] without any prior preprocessing of the FMRI. Techniques from random field theory that, e.g., deal with the underlying multiple testing problem in FMRI (12–14) are thus also applicable in MB estimation. Indeed, whether to proceed the analysis in a massive-univariate voxel-wise manner or by performing a multi-voxel pattern analysis of the test field is not a matter of PB versus MB. However one may proceed, the validity of any analysis can only be guaranteed when the denominator in [5] is correctly estimated. By design, this cannot be the case in PB estimation.

## 5 Basic examples for valid weighting schemes

The weighting scheme that most closely matches prior spatial smoothings by a Gaussian filter in PB estimation is to

[2] Some of the spatial variation will show in form of autoregressive correlation in the voxel-centric time series data. This is not an advantage. The analysis of models with autocorrelated residuals is generally more costly, and the hope to adequately capturing the full spatial variation by the variance inflation factor as estimated by, say, an autoregressive model is doubtful.

define each $\omega^x$ as a translation of a Gaussian kernel function such that it centres at $x$:

$$\omega_r^x = \begin{cases} e^{-\frac{1}{2}\left(\frac{\|r-x\|}{h}\right)^2} & \text{for } r, x \in M, \\ 0 & \text{otherwise.} \end{cases} \qquad [6]$$

Here, $h > 0$ denotes the standard deviation of the respective Gaussian distribution. The parameter $h$ plays the role of a bandwidth parameter. Additionally, weights outside of $M$ were set to 0 in [6], which will produce a discontinuity on the surface of the template brain. This shall reflect that we do not expect to see BOLD signals outside of the boundaries of the brain.

The WLS argument in [3] expects a sufficient density of measurements within the main mass of the weighting scheme $\omega^x$ at $x$ in order to reliably estimate all unknown parameters in [1]. The bandwidth $h$ (or more generally the full width at half maximum (FWHM) of a weighting scheme) is thus basically bounded from below by the spatial resolution of the scanner's acquisition grid.

The scheme [6] is a special case for defining $w_r^x$ as

$$\omega_r^x = \begin{cases} \varphi\left(\frac{\|r-x\|}{h}\right) & \text{for } r, x \in U_x, \\ 0 & \text{otherwise.} \end{cases} \qquad [7]$$

for a smooth kernel $\varphi$ and a compact neighbourhood $U_x \subseteq M$ of $x$. If $\varphi$ has compact support, say $S$, it will speed up the computation, as weights outside of

$$\left\{ r : \varphi\left(\frac{1}{h}\|r-x\|\right) \in S \right\}$$

vanish. If $S$ is a subset of $[0, 1]$, then this area is contained within a ball $B_h(x)$ of radius $h$ around $x$, and weights outside of $B_h(x)$ are guaranteed to vanish. Without loss of generality, it may then be assumed that $U_x \subseteq B_h(x)$.

Construction [7] can further be generalised by replacing $\frac{1}{h}\|r-x\|$ by a divergence map $d(r\|x)$ measuring the discrepancy *of $r$ from $x$*, namely

$$\omega_r^x = \begin{cases} \varphi\left(d(r\|x)\right) & \text{for } r, x \in U, \\ 0 & \text{otherwise.} \end{cases} \qquad [8]$$

The choice of $d(r\|x)$ will typically have a much greater impact on the form and curvature of the fitted parameter fields than the choice of which $\varphi$ to use in the fitting.

## 6 Constructing new weighting schemes from old

One of the modelling assumptions has been that the effect field $\beta_j(x)$ is continuous in $x$ (at least within the brain $M$, i.e. with the exception of surface areas and tissue boundaries). Hence, if $\beta_j(x)$ is the effect at $x \in M$, we expect this effect to only change negligibly in a sufficiently small neighbourhood of $x$. Mathematically this is equivalent to assuming that for any given $\varepsilon > 0$, there exists a neighbourhood $U$ of $x$ for which

$$\sup_{r \in U} \left|\beta_j(r) - \beta_j(x)\right| \leq \varepsilon.$$

Without loss of generality, it can be assumed that $U$ is compact and connected, as $\mathbb{R}^3$ is locally compact. Let $\mathfrak{U}(x)$ be the set of all connected, compact neighbourhoods of $x$, and let $U_\varepsilon(x)$ be the largest such neighbourhood for which the above inequality holds, in signs:

$$U_\varepsilon(x) := \bigcup \left\{ U \in \mathfrak{U}(x) : \sup_{r \in U} \left|\beta_j(r) - \beta_j(x)\right| \leq \varepsilon. \right\}$$

The set $U_\varepsilon(x)$ is one of the (finitely many) connected components of

$$\beta_j^{-1}\left(\overline{B_\varepsilon\left(\beta_j(x)\right)}\right) = \left\{ r : \left|\beta_j(r) - \beta_j(x)\right| \leq \varepsilon \right\},$$

namely the component which contains $x$. Since $\beta_j$ is continuous, this shows that $U_\varepsilon(x) \in \mathfrak{U}(x)$, i.e. that $U_\varepsilon(x)$ is compact and connected. As we shall see, the sets $U_\varepsilon(x), x \in M$ play an important role in the selection, construction and evaluation of weighting schemes.

The inverse direction to the just defined $\varepsilon \mapsto U_\varepsilon(x)$ is defined by

$$\varepsilon_U(x) := \sup_{r \in U} \left|\beta_j(r) - \beta_j(x)\right|$$

for $U \in \mathfrak{U}(x)$. This function $U \mapsto \varepsilon_U(x)$ maps subsets of $\mathbb{R}^3$ to positive scalars. The value $\varepsilon_U(x)$ measures the maximal divergence of the field $\beta_j$ from $\beta_j(x)$ within the set $U$. We shall call $\varepsilon_U(x)$ the *epsilon error* of $U$ in $x$. Note that $U_\varepsilon(x)$ and $\varepsilon_U(x)$ are not actual inverses of each other. $U_\varepsilon(x)$ maps $\varepsilon$ to the largest neighbourhood $U \in \mathfrak{U}(x)$ whose epsilon error is *at most* $\varepsilon$:

$$\varepsilon_{U_\varepsilon(x)}(x) \leq \varepsilon.$$

Also, for any neighbourhood $U \in \mathfrak{U}(x)$:

$$U \subseteq U_{\varepsilon_U(x)}(x).$$

If $x$ is clear from the context, we shall drop the argument $x$ from $U_\varepsilon$ and $\varepsilon_U$.

If $U \in \mathfrak{U}(x)$, then $\varepsilon_U \leq \varepsilon$ if and only if $U \subseteq U_\varepsilon$, since $U_\varepsilon$ is the largest set whose epsilon error is bounded by $\varepsilon$. Both functions $\varepsilon \mapsto U_\varepsilon$ and $U \mapsto \varepsilon_U$ are monotone with respect to $\leq$ and $\subseteq$: it is $\varepsilon_0 \leq \varepsilon$ if and only if $U_{\varepsilon_0} \subseteq U_\varepsilon$. In other words, the sequence $(U_\varepsilon)_{\varepsilon>0}$ is an ascending family of sets. It follows that for any $r \in \mathbb{R}^3$, there exists some $\varepsilon_r > 0$ such that $r \in U_\varepsilon(x)$ for all $\varepsilon > \varepsilon_r$, and

$$\bigcup_{\varepsilon>0} U_\varepsilon = U_\infty = \mathbb{R}^3.$$

This is an important property of the family $(U_\varepsilon)_{\varepsilon>0}$, as it allows the construction of a *native* divergence map for the $\beta_j$-field: for $r \in \mathbb{R}^3$, define

$$d(r\|x) := \inf\{\varepsilon > 0 : r \in U_\varepsilon(x)\}.$$

Then,

$$d(r\|x) = \varepsilon \quad \text{if and only if} \quad r \in \delta U_\varepsilon(x).$$

Conversely, given a divergence map $d = d(r\|x)$, the sets $U_\varepsilon(x)$ can be completely recovered:

$$U_\varepsilon(x) = \left\{r \in \mathbb{R}^3 : d(r\|x) \le \varepsilon\right\}.$$

The family $(U_\varepsilon(x) : \varepsilon > 0, x \in M)$ and equivalently the native divergence map $d$ of $\beta_j$ give a complete description of the to-be-estimated effect field.[3]

If a weighting scheme is explicitly defined as the convolution of a divergence map $d$ with kernel $\varphi$ (see [8]), this transparently shows that $d$ represents a deliberate choice or belief on how BOLD signals spread spatially in an FMRI. Namely, in case of the Gaussian kernel [6], the belief is that the BOLD signal at $x$ spreads in concentric circles via

$$d(r\|x) = \frac{\|r - x\|}{h}.$$

into the neighbourhood of $x$. Any estimate $\hat{\beta}_j$ of $\beta_j$ yields an estimate $\hat{d}$ of the native divergence map of the $\beta_j$-field. If $\omega$ in [3] is replaced by

$$\hat{\omega}^x(r) = \varphi\left(\hat{d}(r\|x)\right),$$

this results in a new estimate, say $\hat{\hat{\beta}}_j$ of $\beta_j$. The procedure may even by iterated: a divergence map $d_n$ yields a weighting scheme $\omega_n$ via $\omega_n^x(r) := \varphi\left(\hat{d}_n(r\|x)\right)$, which will in turn produce an estimate $\hat{\beta}_{jn}$ of $\beta_j$. The cascade is started by an initial choice for $d_0$. The algorithm will produce sequences of divergence maps $(d_n)_{n\in\mathbb{N}}$, weighting schemes $(\omega_n)_{n\in\mathbb{N}}$, and estimates $(\hat{\beta}_{jn})_{n\in\mathbb{N}}$. The final estimate for the $\beta_j$-field is obtained by setting

$$\hat{\beta}_j := \hat{\beta}_{jn_0}$$

for a suitable choice $n_0 \in \mathbb{N}$. The algorithm can further be generalised by choosing $\varphi$ from an a-priori family $(\varphi_n)_{n\in\mathbb{N}}$. This, e.g., allows to choose one kernel $\varphi_0$ for the initial estimate of $\beta_j$ and a different kernel for the rest of the $\varphi_n$, $n > 0$. It demonstrates that inductive procedures arise naturally in the theory of the MB approach to FMRI.

The above iterative algorithm is somewhat different to the algorithm at which Polzehl et al. (9) arrive, which is known as adaptive smoothing or structural adaptive segmentation in FMRI. The two algorithms, however, share similar characteristics and it should be possible to adapt their propagation-separation approach to the here suggested MB estimation procedure. This, though, is outside the scope of this manuscript.

## 7 Power

In the previous section, we have seen that any divergence map $d = d(r\|x)$ represents a specific belief on how the BOLD signal at a point $x \in M$ spreads into the neighbourhood around $x$. Let $d$ denote the native divergence map of the $\beta_j$-field:

$$d(r\|x) = \left|\beta_j(r) - \beta_j(x)\right|, \quad U_\varepsilon(x) = \{r : d(r\|x) \le \varepsilon\}.$$

Since $d$ is unknown, let $\tilde{d}$ denote the divergence map that is used for fitting $\hat{\beta}_j$, i.e. let $\omega^x(r) = \varphi(\tilde{d}(r\|x))$ for some kernel $\varphi$. Let

$$\tilde{U}_\varepsilon(x) = \left\{r : \tilde{d}(r\|x) \le \varepsilon\right\}.$$

Then all observations at the boundary of $\tilde{U}_\varepsilon(x)$ will receive the same weight when estimating $\hat{\beta}_j(x)$. Ideally, though, this should be the case for all points at the boundary of $U_\varepsilon(x)$ instead. As the MR-signal in the neighbourhood around each $x$ is only sampled at a finite grid and up to a limited resolution, power can be gained when the boundaries $\delta\tilde{U}_\varepsilon(x)$ well approximate the boundaries $\delta U_\varepsilon(x)$. (For details see Figure 4.) If $\tilde{d}(r\|x) > d(r\|x)$, bias is introduced into the analysis at $x$. If $\tilde{d}(r\|x) < d(r\|x)$, power is lost.

The selection of the correct weighting scheme gains weight in high resolution FMRI. As the density of observations increases with increasing resolution, $d$ gains more flexibility in the construction of the $\tilde{U}_\varepsilon$, and $U_\varepsilon(x), x \in M$ can be estimated more accurately. The higher the resolution, the more gain in power is to be expected from inductive procedures.

## 8 Bias and mean squared error

Let $\omega$ be a valid weighting scheme and let, without loss of generality, each $\omega^x, x \in M$ be normed, i.e. let us assume that[4]

$$\int_{\mathbb{R}^3} \omega_r^x dr = 1.$$

Let $\hat{\beta}_j(x)$ be defined as in [3]. Then

$$\mathbb{E}\left(\hat{\beta}_j(x)\right) = \int_{\mathbb{R}^3} \omega_r^x \beta_j(r)dr = \int_{\operatorname{supp}\omega^x} \omega_r^x \beta_j(r)dr, \qquad [9]$$

where the integral can be restricted to the support of $\omega^x$:

$$\operatorname{supp}\omega^x := \overline{\{r : \omega_r^x > 0\}}.$$

[3] Divergence maps are a familiar leitmotif in the study of probability distributions. The most famously known example is likely the Kullback–Leibler divergence. Multivariable calculus, though, also knows a concept under the same name. In calculus, divergence is a characteristic of the flow of a vector field. As these two are very different concepts, this manuscript will always speak of the *native divergence map of the scalar field* $\beta_j$ instead of simply referring to $d$ as the *divergence of* $\beta_j$.

[4] Otherwise replace $\omega_r^x$ by $\omega_r^x / \int_{\mathbb{R}^3} \omega_r^x dr$.

If $\omega^x$ has compact support, then $\operatorname{supp}\omega^x \in \mathfrak{U}(x)$. If the integral in [9] is restricted to a subset $V$ of the support, this will produce a deviation to the actual value of $\mathbb{E}\left(\hat{\beta}_j(x)\right)$, namely

$$\delta_V(x) := \left| \int\limits_{\operatorname{supp}\omega^x \setminus V} \omega_r^x \beta_j(r) dr \right|. \qquad [10]$$

For $V \in \mathfrak{U}(x)$, let us call $\delta_V(x)$ the *delta error* of $V$ in $x$. Since $\beta_j$ is continuous in $x$, since $\omega^x$ is positive and $V$ compact, the mean value theorem guarantees the existence of a point $\xi \in V$ such that

$$\beta_j(\xi) = \frac{\int\limits_V \omega_r^x \beta_j(r) dr}{\int\limits_V \omega_r^x dr}. \qquad [11]$$

If it were that $\xi \in U_\varepsilon(x)$, then

$$\left|\beta_j(x) - \beta_j(x_\delta)\right| \leq \varepsilon. \qquad [12]$$

This motivates the following definition:

**Definition** Let $\omega$ be a (valid) weighting scheme and let $\varepsilon > 0$. If there exists some $V \in \mathfrak{U}(x)$ with $\delta_V(x) \leq \varepsilon$ and $V \subseteq U_\varepsilon(x)$, we say that $\omega$ is $\varepsilon$-nice in $x$. We say that $\omega$ is $\varepsilon$-nice, if $\omega$ is $\varepsilon$-nice for all $x$. We say that $\omega$ is nice, if $\omega$ is $\varepsilon$-nice for all $\varepsilon > 0$.

The definition makes a statement about the tails of the $\omega^x$, $x \in M$. Loosely speaking, the tails of the densities $\omega^x$ are not allowed to be too heavy (Figure 5). If the $U_\varepsilon(x)$ are large, the main mass of an $\varepsilon$-nice weighting scheme is allowed to extend to a certain degree into the neighbourhood of $x$, thus increasing the power of the analysis. But if the sets $U_\varepsilon(x)$ are narrow, weights in $\omega^x$ need to vanish quickly as this would otherwise lead to biased estimate of the BOLD effect at $x$.

If $\omega$ is not nice in $x$, [12] must no longer hold, as any $\xi$ in [11] may lie well outside of $U_\varepsilon(x)$. The following theorem shows that nice weighting schemes produce unbiased estimators, and the bias introduced by choosing a weighting scheme that is not nice in $x$ can be quantified.

**Theorem** *If $\omega$ is $\varepsilon$-nice in $x$, then the bias of $\hat{\beta}_j(x)$ for $\beta_j(x)$ is bounded by $2\varepsilon$. If $\omega$ is nice in $x$, then $\hat{\beta}_j(x)$ is an unbiased estimator of $\beta_j(x)$.*

*Proof.* For $\varepsilon > 0$, let $V \in \mathfrak{U}_\varepsilon(x)$ be such that $\delta_V \leq \varepsilon$ and $V \subseteq U_\varepsilon(x)$. Let $\xi \in V$ be such that [11] holds for $\xi$ and $V$. Then

$$\begin{aligned}\left|\mathbb{E}\left(\hat{\beta}_j(x)\right) - \beta_j(x)\right| &= \left|\mathbb{E}\left(\hat{\beta}_j(x)\right) - \beta_j(\xi) + \beta_j(\xi) - \beta_j(x)\right| \\ &\leq \left|\mathbb{E}\left(\hat{\beta}_j(x)\right) - \beta_j(\xi)\right| + \left|\beta_j(\xi) - \beta_j(x)\right| \\ &\leq \delta_V + \varepsilon \leq 2\varepsilon.\end{aligned}$$

If $\omega$ is nice in $x$, then $\varepsilon$ can be chosen arbitrarily. □

Let

$$\varepsilon_\omega(x) = \inf\{\varepsilon > 0 : \omega \text{ is } \varepsilon\text{-nice in } x\}.$$

Note that $\varepsilon_\omega(x)$ is well defined, since if $\omega$ is $\varepsilon$-nice, also $\omega$ is $\varepsilon'$-nice for all $\varepsilon' > \varepsilon$. If $\omega$ is nice in $x$, then $\varepsilon_\omega(x) = 0$. A direct corollary of the above theorem is that

$$\left|\operatorname{Bias}\left(\hat{\beta}_j(x), \beta_j(x)\right)\right| \leq 2\varepsilon_\omega(x).$$

Consequently, a bound for the mean square error of $\hat{\beta}_j(x)$ is

$$\mathbb{E}\left(\left(\hat{\beta}_j(x) - \beta_j(x)\right)^2\right) \leq \sigma^2\left(\hat{\beta}_j(x)\right) + 4\varepsilon_\omega^2(x).$$

Equipped with $\supsetneq$, the family $\mathfrak{U}(x)$ is a net, and it immediately follows that $\xi_V \to x$ for $V \in \mathfrak{U}(x)$. Since $\beta_j$ is continuous,

$$\beta_j(x) = \lim_{V \in \mathfrak{U}(x)} \beta_j\left(\xi_V\right).$$

This continuous statement holds, indeed, for all weighting schemes. The crux with $\varepsilon$-nice schemes is that the corresponding delta-errors $\delta_V$ are bounded from above by $\varepsilon$: $\delta_V \leq \varepsilon$ for all $V \in \mathfrak{U}(x)$.

## 9 Validity

The MB method estimates the variance of the spatio-temporal error terms by evaluating the projection of the (unaltered) FMRI onto the regression surface of the model. This has the consequence that the estimator is able to yield an actual estimate of

$$\sigma^2\left(\hat{\beta}_j(x)\right).$$

This appears to have the intriguing consequence that the MB estimator is able to control its type I error even under not quite optimal choices for the weighting schemes that where used during the fit. See Figure 9 for an illustration of this aspect. Of course, poorly chosen weighting schemes increase the variance of all estimates, and this will negatively impact the power of the overall analysis. When the FWHM of a weighting schemes $\omega$ is sufficiently large, though, the MB estimator appears to have the ability to protect the analysis from moderate misspecifications of $\omega^x$ in each $x \in M$. It is because the bias that is introduced into the analysis by $\omega$ also increase the variance of the error terms at $x$. The larger the bias that is introduced by a wrongly chosen weighting scheme at a point $x$, the lager will be the residual variance of the model at this point, and the less likely an effect at $x$ will be reported as significant.

## 10 Fitting the population model

It is biologically and physically plausible to model indi-vidual, subject specific activations at a point $x \in M$ as a random effects model with two kinds of variance components: a heterogeneity parameter that models the deviation of the expected activation field of an individual subject from the population average, and heteroscedasticity parameters modelling the deviation of the actual observed subject activation field from its expectation due to uncontrollable external variables. If $\varepsilon_{jt}$ and $\xi_j$ are distributed according to

$$\varepsilon_{jt}(x) \sim N\big(0, \sigma_j^2(x)\big), \quad \xi_j(x) \sim N\big(0, \tau^2(x)\big)$$

at every $j, t$, and $x$, this yields the following model for the expected activation $\gamma(x) + z_j^\top \delta(x)$ at $x \in M$ in the population:

$$\hat{\beta}_j(x)\Big|_{\beta_j(x)} \sim N\left(\beta_j, \sigma^2\left(\hat{\beta}_j(x)\Big|_{\beta_j(x)}\right)\right),$$
$$\beta_j(x) \sim N\left(\gamma(x) + z_j^\top \delta(x), \tau^2(x)\right),$$

where

$$\sigma^2\left(\hat{\beta}_j(x)\Big|_{\beta_j(x)}\right)$$

denotes the variance of the estimator $\hat{\beta}_j$ given $\beta_j$. This model is called a random effects meta regression model. It was first introduced by Berkey et al. (15) to perform the meta analysis of clinical trials while accounting for cohort specific covariates. Since then, the model has been studied extensively, various estimates for the heterogeneity parameter have been proposed (16–20), and various test statistics (21, 22) have been suggested, which allow valid statistical tests on the covariate parameters of the model. Since canonical test statistics for the covariate components have inflated type I error (21, 22), an adjustment by Knapp and Hartung (22) is now seen as the state-of-the-art when performing inference in this model. The adjustment multiplies the variance estimate in the denominator of the canonical t-test statistic by a profile function for the heterogeneity parameter.

Poline and Brett (1) point out that taking into account the variances involved when estimating individual activation fields and adjusting their contribution to the overall population effect accordingly increases the accuracy of the latter. But, again, this is not the most important vantage point here. Failing to consider these variances inevitably inflate the type I error of an analysis. As before, the crux of the here presented estimators lies in the possibility to accurately estimate their variances. Only then can valid statistical conclusions be drawn.

An estimator for the heterogeneity that works well with the Knapp–Hartung adjustment is the Hedge estimator (17). In contrast, e.g., to the still widely used restricted maximum likelihood (REML) estimator for heterogeneity estimation, the resulting covariate estimators do not show an inflated type I error. Hedge's estimator has two additional advantages here: being a method of moments estimator, (i) the calculation is computationally feasible in the FMRI setting, and (ii) the estimator makes no distributional assumptions and yields consistent estimates even under deviation from normality.

## 11 Fitting grey scale heterogeneity

FMRI images are typically exported, saved, and distributed in various formats such as DICOM or Nifti. In order to tailor the grey scale values of the FMRI to the integral data type of the respective format specifications, grey scale intensities are shifted and scaled accordingly (see Figure 6). Other biomedical imaging techniques are able to report intensities in specific units, e.g., computed tomography (CT) scans may use the Hounsfield scale for their images. In contrast, MR images have no unit. This is not an issue as long as only qualitative image characteristics are of interest, as they are typically extracted from structural T1-weighted MRI scans. In FMRI studies, though, key interest lies in the quantitative changes of the MR signal, namely the BOLD contrast modelled as the $\beta_j$ covariate in [1].

The ratio $\beta_j(x)/\alpha_j(x)$ is invariant with respect to image scaling, and indeed the mean signal intensity $\alpha_j$ at $x$ serves as the reference whether to consider $\beta_j(x)$ as large or small in magnitude. In order to compare BOLD estimates across subjects and across FMRI sessions, this reference has to be fixed to a common value, say $\mu_x$.

If $c_j\beta_j(x)$ is the BOLD effect that is to be expected with respect to the reference $\mu_x$, this modifies the population model to

$$c_j\beta_j(x) \sim N\left(\gamma(x) + z_j^\top \delta(x), \tau^2(x)\right).$$

It must be $c_j(x) = \mu_x/\alpha_j(x)$ as only then

$$\frac{c_j\beta_j(x)}{\mu_x} = \frac{c_j(x)\beta_j(x)}{\mu_x} = \frac{\frac{\mu_x}{\alpha_j(x)}\beta_j(x)}{\mu_x} = \frac{\beta_j(x)}{\alpha_j(x)}.$$

Any regression surface is a smooth version of the »truth« with model dependent curvature. This is particularly also the case for the estimated intercept fields $\hat{\alpha}_j$ in a FMRI study. It is thus sensible to chose a reference field $\mu_x$ with similar curvature. As the curvature of any $\hat{\alpha}_j$ is predominately influenced by the choice of the weighting scheme in [3], a canonical choice for the reference field is to smooth the template $M$ by the same kernel. If this is the case, we say that $c_j\beta_j(x)$ is the BOLD signal at *$x$ above template intensity* (ati).

Any heterogeneity of the grey scales in a sample of FMRIs reduces the power to detect task related BOLD signals. If the heterogeneity is small, then so will be the power loss. The power reduction becomes an issue for the analysis whenever differences at a point $x \in M$ that are due to the differences in scales are larger in magnitude than the respective BOLD signals at this point. As long as the heterogeneity in a sample is not systematic, an impact on the type I error of a study can fortunately be ruled out. This also means that norming FMRI to ati will by indifferent to the type I error of a study. The norming will, though, increase the power of a study as it effectively reduces grey scale heterogeneities to zero across a sample. This is of substantive relevance in multicentre studies where subjects are measured on different MR scanners, over a longer period of time, and potentially varying acquisition parameters, i.e. whenever the grey scale heterogeneity in a data set is expected to be large.

A side effect of norming data to *ati* is that the norming enables analysts to also study, report, and compare effect sizes of distinct FMRI studies.

## 12 Implementation

All methods have been implemented in a software tool available at `https://fmristats.github.io/`. The software is open source, free to use, and free to distribute under the GPLv3 licence.

## 13 Model building and selection

The different view on FMRI data, which stands at the basis of MB estimation, has consequences on the model building and model selection process.

FMRI data have a spatio-temporal correlation structure, which is typically studied in the form of time series data at voxel-level, i.e. in data which have seen at least some motion correction; the concept of a »voxel« really only exists after motion correction. Without motion correction, one can only sensibly talk about points $x$ in a subject and measurements which have been taken in the vicinity of $x$. Voxel-level data are always the result of a functional transformation of the original data, and the author is unaware whether the unaltered spatio-temporal series of MR signals – the data which actually contribute to the intercept and effect estimates at a point $x$ – have ever been depicted in the literature before.

For a two-block task design, Figure 7 shows such a plot. All data shown in the Figures **7**, **8**, and **9** show data from the first subject that was recruited in a sample of 64 right-handed, healthy subjects that were part of a study that focused on language processing (see the materials section for details). Figures **10**, **11**, **12**, **13**, and **14** are based on the data of the complete sample of 64 subjects.

Figure 7 shows all MR measures of this first subject during task blocks within a radius of 6.90 mm around a point $x_0 \in M$. Let us denote the subject by $j_0$ for the remainder. Depicted are 3,013 observations scaled in size with respect to their distance to $\psi_{j_0}(x_0)$. The smaller a dot, the larger its distance to $\psi_{j_0}(x_0)$. The plot approximately shows all data used by the PB estimator for effect estimation at $x_0$ when data are spatially smoothed via a Gaussian filter with FWHM $=$ 5.42 mm prior to effect estimation.[5] The plot shows exactly the data used by the MB estimator when using a Gaussian weighting scheme with the same kernel.[6] The mean signal per block is shown as a solid line.

First, notice the relatively large signal variance around $x_0$ in relation to the relatively low temporal variation along the time axis. As the PB estimator spatially smooths the data prior to the statistical analysis, this spatial variation is hidden from downstream statistical tests. The resulting overconfidence for the estimated effect at $x_0$ empirically shows as an inflated false positive rate in FMRI studies, e.g. (4,10, 11). Furthermore, inference in PB estimation is only based on an aggregated sample with one value per scan cycle. Here, the PB estimator would only use 94 aggregated values for fitting its underlying GLM at $x_0$. This sample size reduction is drastic. The MB estimator at the same point uses mean (median) observations of 10.8 (16) per slice and time point, 32.0 (32) observations per scan cycle, 188.3 (192) per block, and in total 3,013 observations for inference at $x_0$.

Second, task related effects, i.e. mean signal differences between blocks, are visual, but otherwise measurements appear time stable, in particular within task blocks (a formal test confirms this impression, see below).

Third, effect sizes seem to vary with the position of the respective block in the task sequence. When the stimulus block is presented the first time (see $A_1$), the mean signal increase within this block is the largest across the entire session. When $A$ is repeated without an intermediate control block (see $A_2$ to $A_3$ and $A_7$ to $A_8$), the effect is reduced. Apart from these two reductions and the strong response at the beginning, mean signals in $A$ appear stable across the stimulus blocks. Mean signals in $B$, though, appear to gently increase over time: $B_1, B_2, B_3$ are below, $B_4, B_5, B_7$ are at, and $B_6, B_8$ are above the respective task mean. When going from $B_7$ to $B_8$ without an intermediate $A$ block, the mean signal is elevated. This suggests that block effects possess a rather complex entangled structure in FMRI.

Forth, no residual autocorrelation is visible in the scatter plot. This is somewhat surprising, as there exists an extensive literature on how to deal with autocorrelation in FMRI voxel-based time series. We shall therefore come back to the issue of residual autocorrelation in MB estimation in the next section.

Three model designs were fitted to the data: (i) a nested, two-factorial design with categorical factors *task* and *block* in which block is nested in task, (ii) a two-factorial design with the factor *task* and a linear term for *time*, and (iii) a polynomial b-spline model with the categorical factor *task* and 12

[5] Which observations are precisely used by the PB estimator for effect estimation depends on the individual set-up of the preprocessing steps that are part of the PB approach.

[6] The FWHM = 5.42 mm of a Gaussian kernel equals a standard deviation of $\sigma = 2.30$ mm of the respective distribution. Hence, 99.7% of the kernels tangible mass are contained in a ball of radius $3\sigma = 6.90$ mm. To speed up computations, observations further than $3\sigma$ from $x_0$ were set to 0.

degrees of freedom for *time*.

All models were fitted on a common $(2 \text{ mm})^3$ grid which covered the brain of the respective subject. The highest elevation in each of the respective t-statistic fields (which test for non-zero task related effects) lies at the *same* point, namely $x_0$. A heat map of a transversal cut through $x_0$ is shown in Figure 8. The task effect at $x_0$ was estimated to 15.0, 15.0, and 14.8 in the respective models (i), (ii), and (iii) with respective t-values 7.68, 7.57, and 4.90. These correspond to (nominal) p-values of $<3 \cdot 10^{-14}$, $<5 \cdot 10^{-14}$, and $<2 \cdot 10^{-6}$. A test on time as a significant covariate in model (ii) has a p-value of 0.7734. An omnibus test on the significance of any of the b-spline coefficients in (iii) has a p-value of 0.0239.

Model (i) is a parsimonious way to model entangled interactions between blocks, as it directly models the nested design of the data. The model is fully saturated, i.e. mean signal intensities are separately modelled for each block. The model is parametric in the task/block-dimension but non-parametric in time: the model fits a step function to the temporal course of the signal and it will follow temporal drifts and fluctuations with unmatched flexibility.

This modelling strategy, however, is rather unusual in the FMRI literature.[7] It is more common to anticipate – by model selection criteria or by an a priori choice – the shape of the signal in- and decrease of the BOLD contrast, to fold this function with $\mathbb{1}_t$, to assume that task related changes are constant across the blocks in the experiment, and to add numerous covariates to the model which allow the model to follow temporal fluctuations (1, 2). Added covariates may include time in multiple polynomial degrees, low frequency cosine functions, or aggregated variables such as head movement parameters in multiple degrees (1, 2). As many of these variates will have only negligible effects on the signal course, their inclusion inflate the conditioning number of the underlying regression problem.

Two statistical consequences of the latter strategy can be illustrated by models (ii) and (iii). As the task related signal response is assumed to be constant in (ii), the model is unable to follow the individual block deviations from the respective common mean (see blocks $A_1$, $A_3$, $B_6$, $B_8$, and $A_8$ in Figure 7). Consequently, the model has to credit these deviations to apparent signal fluctuations. This is nicely seen by the fit of model (iii). Also, model (iii) is unable to attribute variable effect strengths to individual task blocks, but its high temporal flexibility allows the model to nevertheless approximate the respective mean signals in each block, i.e. to approximate the fit of the fully saturated, nested model (i). The increased model fit, though, is bought at the price of overfitting. The consequence is a loss in power – quite noticeable by the drop from 7.10 to 4.42 of the respective t-value for the task effect.

Neither is nor can this be systematic treatment of the complex topic of model building, selection, and validation. It shall highlight, though, the transparent access of the MB estimator to the data at every single point. Current methods are only able to study FMRI data on aggregated values, as they are forced to reduce the data to voxel-level first. This original transparency will enable new, in-depth studies of the spatio-temporal structure of FMRI data.

## 14 Autocorrelation

The scatter plot in Figure 7 does not suggest the existence of autocorrelation of any considerable impact. This is somewhat surprising, as considerable effort exists in modelling and fitting residual autocorrelation in FMRI voxel-based time series data.

A major cause for the presence of autocorrelation is typically the omission of one or several key variables from a model (23). If the omitted variables positively correlate with time, error terms in a model will show positive autocorrelation, since they still include the effects of the missing variable. This is important: residual autocorrelation does not reflect the correlation structure of data per se, but it is a sign for how much *leftover* correlation is still present in the data that could not be explained by the model. Residual autocorrelation hints on the existence of explainable variables that are currently missing from deterministic parts of the model. If it is possible to include these variables in the model, their inclusion will increase the validity and power of an analysis. If this is not possible, in the case where these variables are unobservable, one must revert to modelling the autocorrelated error structure in order to avoid underestimating the variance of the error terms.

Mathematically speaking, the PB estimator prepares data by a functional: the estimator does not fit a model to the original FMRI, say $y$, but to $F(y)$, and the functional $F$ depends on various parameters including subject movement. At the end – and indeed already after motion correction – each voxel intensity in the final image is a function of measurements in the spatio-temporal proximity of the respective voxel. Prepossessing an FMRI introduces a functional spatio-temporal dependence structure. It should be expected to see residual autocorrelation when fitting a model to a preprocessed FMRI.

As a consequence, complex models have to be used to model the resulting error distributions in the PB approach. Fitting a regression model with autocorrelated error structure is not only statistically more complex but also computationally expensive. Also, the kind of autocorrelation that should be allowed becomes an issue (24). In practice, it is common to model the error distribution via an AR(2) process (1, 2); not necessarily because it is known to fit the correlation structure best but it is the most parsimonious error model which is still computationally feasible (2). Misspecifications of the residual structure increase the type I error and decrease power.

[7] The author has not found any applications of a model with nested effects to FMRI data in the literature.

A measure to assess residual autocorrelation in time series is the Durbin-Watson (DW) statistic (25). The statistic is typically applied to FMRI data which have seen (at least) some sort of motion correction, and is then calculated for the respective time series data at the voxels of the »corrected« 4D picture. As voxel-level data are always the result of at least one functional transformation, the presence of residual autocorrelation should be expected here, and the typically small values of the respective DW statistics suggest that autocorrelation is a relevant issue for voxel-based estimators.

The transparent access of the MB estimator to the FMRI data allows to study the full spatio-temporal residual structure of the unaltered time series, i.e. of the »uncorrected« FMRI. The DW statistic takes values between 0 and 4, and the statistic is approximately equal to $2(1-\rho)$ when $\rho$ equals the residual autocorrelation of the model. Values close to 0 suggest positive, values close to 4 negative, and all values around 2 suggest no residual autocorrelation. The DW of the noise in Figure 7 is 1.91.

A poor model fit at points $x$ at which a method fails to recognise significant effects reduces power. In contrast, poor fits in high peak regions, i.e. at points which are claimed as significant by the model, inflate the type I error of a study. In particular the latter has to be avoided as it challenges the validity of a study. Thus notably in high peak regions of the $t_j$-field of a subject, the DW distribution should not suggest the presence of residual autocorrelation. If positive autoregressive correlation exists, the DW distribution should show a clear shift towards 0 while leaning distinctively away from 2. Figure 10 shows the density of the empirical DW distribution in the fits of the nested model to the FMRI of 64 different subjects. The figure shows an estimate of the whole-brain DW distribution and estimates of the same within the high peak regions of the respective brains. Each density estimate is based on all $\mathrm{DW}_j(x), x \in M$ for which $t_j(x) > \hat{t}_{j,\gamma}$ where $\hat{t}_{j,\theta}$ denotes the $\theta$-quantile of the empirical $t_j$-distribution of the $j$th subject. Shown are DW density estimates for the whole brain ($\theta = 0$) and the peak areas $\theta = 0.99$, $\theta = 0.999$, and $\theta = 0.9999$ respectively.

The more $\theta$ approaches 1, the more important it is for the empirical DW distributions not to show any signs of deviations away from the centre at 2. The contrary is the case. All distributions are well centred around 2 and spread symmetrically to each side. This is not the DW distribution of an autocorrelated stochastic process.

## 15 Inference

The subject analysis of subject $j_0$ shown in Figure 8 suggests that there is increased activation in the occipital lobe when performing a word generation task. Let us study whether this finding can be verified in a sample of 31 other right handed subjects. For this, a meta analysis is performed at the suggested point $x_0$ in $M$ that excludes $j_0$. The result of the meta analysis is shown in Figure 11. The figure shows two popular plots that are commonly used to visualise the results of a meta analysis: a forest plot and funnel plot.

Forest plots show the point estimates and confidence intervals for the respective effect for each subject in the study. The effects are typically ordered on the y-axes with respect to some covariate, here a score for handedness. The population inferred effect is usually displayed at the bottom. A funnel plot shows the same effects but plotted against the reciprocal of the standard error of the estimate on the y-axes. On population level, the estimated BOLD effect at $x_0$ is 5.97 ati with a 95%-confidence interval of $[2.29\text{ ati}, 9.66\text{ ati}]$. The p-value for the effect at $x_0$ is 0.0044. The confidence interval and the p-value for the population inferred effect are Knapp-Hartung adjusted. Since the population model was fitted using a sample that excluded the initial subject and since $x_0$ had been chosen a-priori for this analysis, neither the confidence interval nor the p-value have to be adjusted for multiple testing. We may reject the null hypothesis of no BOLD signal at $x_0$ in the population with 1% significance. Figure 12 shows a heat map of transversal cut through the field with the location $x_0$ highlighted by $\oplus$.

In an explorative setting, the population model is fitted to the complete sample of 64 subjects (including the initial subject). The highest peak $x_1$ in the Knapp-Hartung adjusted t-statistics field has been selected. The forest and the funnel plot of the analysis at $x_1$ are shown in Figure 13. Figure 14 shows a heat map of transversal cut through the field with the location $x_1$ marked by $\otimes$. Again, the population inferred task effect is displayed at the bottom in the forest plot. At $x_1$, we may expect an average BOLD effect in the population of 19.99 ati with a 95% confidence interval of $[17.12\text{ ati}, 22.87\text{ ati}]$. The Knapp-Hartung adjusted t-statistic is 11.61 at this point. A t-value of 11.61 corresponds to a p-value below double-precision floating-points. As the point $x_1$ had been chosen posteriori, though, the test would need to be adjusted for multiple testing. As the test field is smooth, a possibility is e.g. to use methods from random field theory to deal with the multiple testing problem (13, 14,26,27).

## 16 Discussion

Both discussed approaches, the standard picture based (PB) estimator and the newly derived model based (MB) estimator, provide smooth effect estimates for task related BOLD contrast in an FMRI – in this sense, both approaches »smooth« the observed data. PB estimation involves two separate smoothing steps: first a prior spatial smooth (that involves data corrections, movement corrections, slice timing corrections, normalisations, and the application of spatial filters) followed by a second, model driven temporal smooth (when fitting the univariate models to the respective voxel-based time series data). The MB approach com-bines all these steps into a single fit of a statistical model.

Typically three arguments are brought forth in order advise and justify prior smoothings of FMRI. Data are smoothed (i) to accommodate for inter-subject variation in brain anatomy, (ii) to increase the signal to noise ratio (SNR) of the data, and (iii) to enable inference from random field theory. Let us discuss each of these arguments separately.

*Argument (i):* The choice of the kernel in MB estimation runs parallel to the same choice that has to be made in PB estimation. Both approaches provide spatially smooth parameter fields as well as smooth test statistic fields. Either approach is equally able to accommodate for inter-subject variation. Both approaches, MB and PB alike, allow to recursively construct weighting schemes in order to increase the power for detecting true, task related BOLD effects in an FMRI. As the spatial smooth is already part of the fitting procedure of the MB approach, prior smoothings are obsolete in MB estimation.

*Argument (ii):* The variance reducing effects of each of the PB preprocessing steps must not be seen as beneficial but as a fundamental flaw of the method. The mathematics is crystal clear in this respect: A test statistic relates the signal in data to the noise in the data. Reducing the noise, e.g. by interpolations of imaging data onto new grids and then dropping the original data (a.k.a. registration) or by actively smoothing data, which mere purpose is to enhance the SNR, decreases the denominator in a test statistic. As the denominator decreases, the test statistic itself is artificially inflated at the same time. When testing for an effect now, the data will appear less likely under the respective test statistic's null hypothesis than they actually are. Ergo, the likelihood to falsely reject the null is inflated. Beautifying the SNR of data shakes at primary principles of statistical decision making.

Smoothing is not bad per se. One perfectly good reason to smooth data is to prepare data, e.g., for visualisation. As humans are not very good in coping with low signal to noise ratios, and, as smoothing filters typically increase this ratio, it allows the analyst to scan the data for effects, detect potential confounding factors, outliers, artefacts, and to assess the general form of the data. Specifically in biomedical imaging, smoothing plays an important role in the preparation of data for visual representations which can reveal internal and hidden structures, aid in diagnosis, or help planing surgery. Whenever the intention is to *look* at imaging data, smoothing techniques can be helpful.

Despite being a vital tool in data visualisation, one needs to be aware that preprocessing data is statistically highly problematic. Whenever the intention is not to look at imaging data in the form of a visual representation but to treat imaging data *as data*, as it is the case in a statistical analysis, prior smoothings (and indeed any technique which increases the SNR whether by intention or not) has the very likely potential to falsify the complete analysis.

The SNR increase of SNR enhancing techniques is in general bought at the price of an overall loss in information. Thus beside the inflation in type I error, smoothing also reduces statistical power. Munafo et al. (28) argue that a field that is chronically working in an underpowered environment is at continuous risk of failing to recognise p-value hacking and data dredging, as confirmation biases may encourage the acceptance of beneficial outcomes, and, as their prime example, the authors list FMRI.

As the MB method estimates the variance of the spatio-temporal error terms by evaluating the complete projection of the unaltered FMRI onto the regression surface of the model, the estimator is able to yield an actual estimate of

$$\sigma^2\left(\hat{\beta}_j(x)\right).$$

This has the stunning consequence that the MB estimator is able to adequately control its type I error even under sub-optimal choices for $\omega$. This is quite astonishing: Poorly chosen weighting schemes, of course, increase the variance of all estimates, and this will negatively impact the power of an analysis. However, when the FWHM of a weighting schemes $\omega^x$, $x \in M$ is sufficiently large, the MB estimator appears to have the ability to protect itself from moderate misspecifications of $\omega^x$ in $x$. The reason is that bias that is introduced into the analysis by $\omega$ also increase the signal variance at $x$ (Figure 9).

*Argument (iii):* Techniques from random field theory, which test for the significance of a peak in a random field, require test statistic fields to be spatially smooth. Since the error distribution in an FMRI is not smooth, the univariate nature of the PB approach virtually forces to spatially smooth the data prior to the statistical analysis. Without spatial smoothings, the PB estimator cannot guarantee the smoothness of any parameter or test statistic field. In contrast, the MB estimator provides smooth parameter fields without any additional assumptions about the error distribution of an FMRI.

There exists a strong belief that without preprocessing, a statistical analysis of FMRI data is invalid (2). The here derived estimators, and in particular their reasoning, show that the contrary is indeed the case.

Despite the false positive inflation (4), despite the ability to show significant activations in 90.3% of all brain voxels when only applying enough different pipelines to FMRI data (11), and despite the delicate difficulties in reproducibility of FMRI results (10,11,28), the standard analysis approach to FMRI data has not yet diverted from the basic voxel-centric approach of the PB estimator. This is likely the case due to the simple lack of feasible alternatives. Preprocessing techniques are widely applied in the field, they are treated as unavoidable and without alternative, and statistical methods are routinely supplied with beautified data. The MB estimator aims to provide this alternative.

The MB approach enables trustworthy access to the vari-ance in BOLD estimation, and it allows analysis of FMRI data without altering original observations. Sound variance estimates are the key ingredient in a statistical analysis. Only with plausible variance estimates is a statistical test able to accurately estimate the risk for reporting a false positive result. Potential effects are not smoothed into the background noise, neither by interpolations nor by applying prior smoothings of any kind. This will increase the power of FMRI studies. Instead of applying any »corrections« to the data, movement information and slice timing differences directly enter the MB estimator. Technical and accessible criteria have been formulated which assure smoothness of activation field estimates and test statistic fields, allowing application of established random field techniques.

The ability to estimate the conditional, local variance of the signal field enables the application of random effects meta regression techniques in the context of FMRI. It has been shown how these models are able to properly model the experimental design of FMRI studies. Meta regression models allow the uncertainty in the estimated individual BOLD effects to pass on to the model used for population inference. Methodological advances in meta analysis of the past 20 years are now available to FMRI.

When analysing data from multicentre studies, in which subjects have been measured on different MR scanners, over a longer period of time, and with potentially varying acquisition parameters, norming to *ati* will increase the power of a study as it effectively reduces the heterogeneity of the mean signal fields to zero across the study and across all $x \in M$. Apart from an increase in power and interpretability, reporting effect sizes in a common unit allows analysts to not only solely look at statistical significance, but to report, to study, and to compare effect sizes of FMRI studies, too.

Beside the apparent differences between MB and PB estimation, the former offers the field of cognitive neuroscience the opportunity to be able to treat FMRI data *as data* rather than as mere series of images. Statistically, MB estimation will enable new methodological advances, as it offers a genuine and statistical transparent framework for model building, model diagnostics, and model selection.

## 17 Materials

FMRI data of 64 subjects were sampled on a 3 T MRI scanner (Tim Trio, Siemens, Erlangen, Germany) equipped with a 12-channel head matrix Rx-coil via a T2*-weighted gradient echo planar imaging (EPI) sequence (TE 25 ms, TR 1450 ms, slice thickness 4.6 mm, FoV 192 mm, flip angle 90° with 30 axial slices oriented perpendicular to the inferior-superior-axis, and 64×64 in-plane resolution). Stimuli were organised in a two block design. Block $A$: silently find as many words as possible starting with a displayed letter. Block $B$: silent repeat of the word »baba«. Block durations: 15 s. Each block type was presented eight times in pseudo-randomised order. All subjects gave their written informed consent prior to participation.

FMRI data were converted from DICOM to Nifti1 using the tool `dcm2nii`. Import and export of Nifti1 files to and from python were handled by `nibabel` (29). A foreground/background separation was performed using Otsu's method (30) via the tool `nii2session` which is part of the `fmristats` tool box. Head movements were estimated by `fmririgid` (also `fmristats`) via a principles component approach. The tool also includes an algorithm based on Grubbs' test for outlying observations (31) that marks scan cycles with suggested severe head movements as potentially problematic. Data from such scan cycles were excluded from the study. Maps from MNI standard space to the respective subject brains in the study were estimated using the tool `RegistrationSyNQuick` which is part of the ANTS tool box (32). The template `MNI152_T1_2mm_brain.nii.gz` that is provided by FSL has been used as a reference for MNI standard space. For each subject, the $42^{nd}$ scan cycle was used as a reference for the subject space if this scan cycle was not marked as an outlier by `fmririgid`. The $84^{th}$ scan cycle was used as a fall back. The actual fit of the signal models were performed with `fmrifit` (part of `fmristats`). Potential non-brain areas were pruned from the estimated parameter fields using `fsl4prune` which contains a wrapper to FSL's `bet`. The program `atlasquery` (also part of the FSL) has been used to query the location of points of interest in MNI standard space. The initial subject $j_0$ in the analysis of Figures **7**, **8** and **9** is the subject which had the least ID in the sample. The meta analysis was performed by using `fmristats`'s python interface. All code of the analysis can be found on `fmristats`' website:

`https://fmristats.github.io/`

## Acknowledgements

I thank Astrid Dempfle and Andreas Jansen for discussions and valuable feedback on this work. The research was funded by the German Research Foundation (Grant No. FOR 2107, DE 1614/3-1).

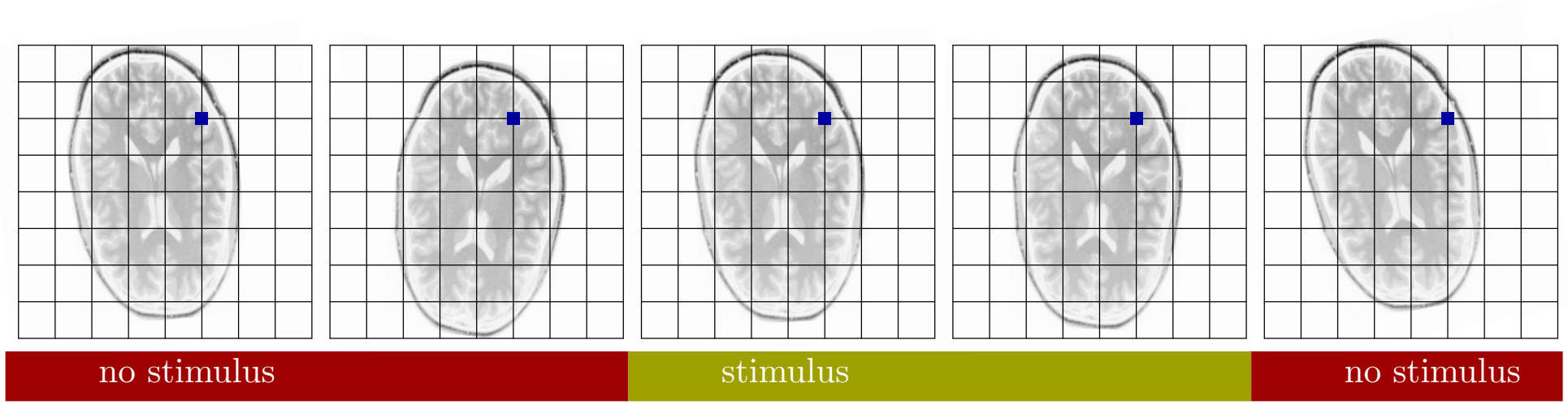

(i) The voxel-centric view of FMRI data: A moving object behind a fixed grid.

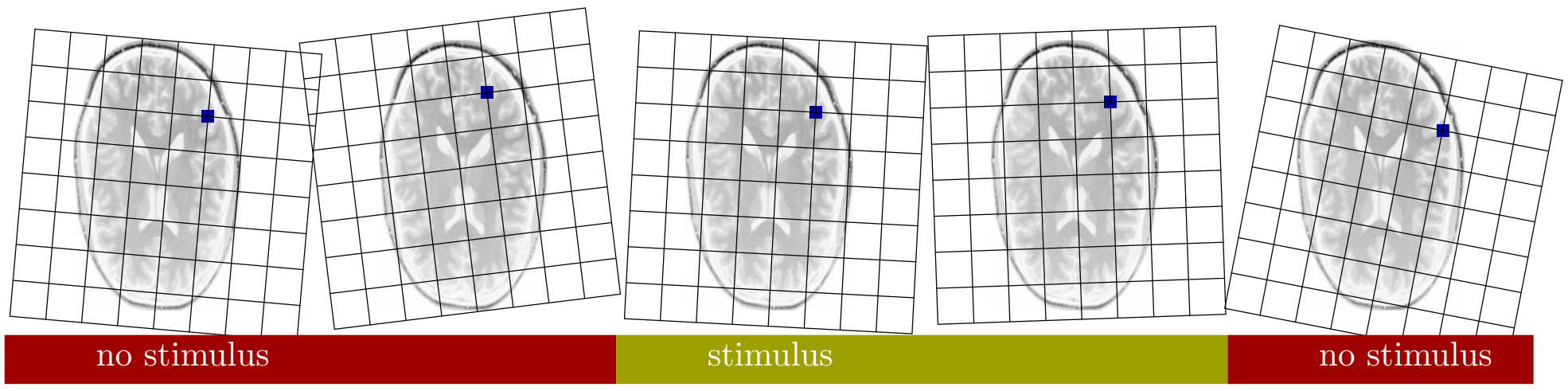

(ii) The data-based view of FMRI data: A moving grid above a fixed object.

**Figure 1:** Different ways of looking at same data. (i) Voxel-centric approaches treat FMRI data as pictures that live on a fixed 3D imaging grid. The analysis of the data is made complicated by movements of the object of interest. (ii) This manuscript propagates a data-based approach to FMRI. The method treats imaging data not as images but as a discrete samples of measurements taken at varying locations in space and time.

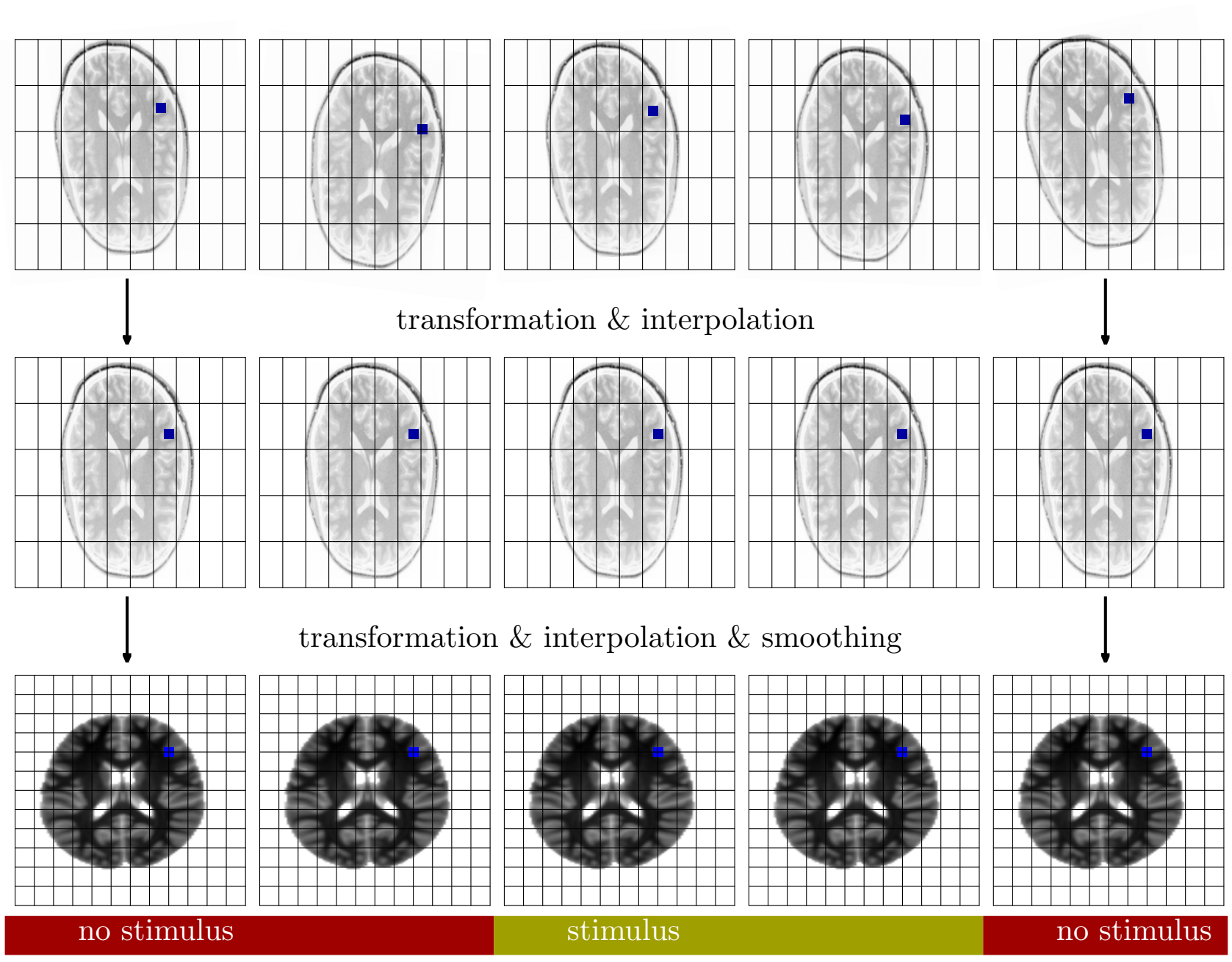

**Figure 2:** Voxel-centric, picture based approach to FMRI analysis. The original FMRI are corrected for subject movements and slice timing differences and interpolated onto a common grid in a standard space. One voxel $v$ in standard space is exemplarily emphasised by a blue square in the figure as well as the voxel's respective location in the subject. The coordinates of $v$ in the scanner varies with the movements of the subject, and so does the interpolation base for the estimated intensity at $v$ in the bottom images. The signal to noise ratio of the data increases from top to bottom.

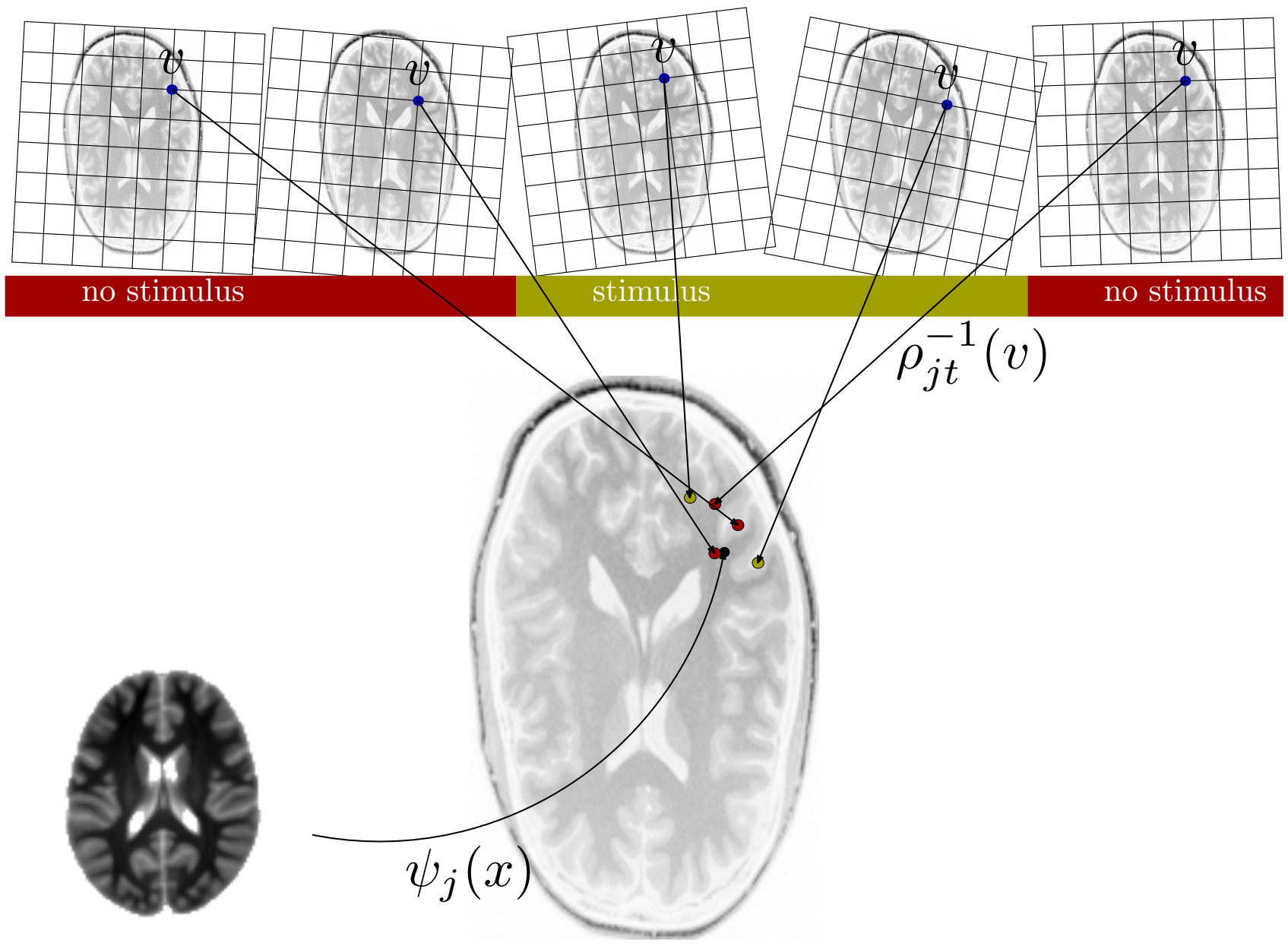

**Figure 3:** Model based approach to FMRI analysis. Samples of the MR signal are taken at various different points in space and time. Maps $\rho_{jt}$ provide information at which points any specific subject $j$ is sampled during $t$. Diffeomorphisms $\psi_j$ give information about the position of a point $x$ in standard space in the subject. Models at $x$ are fitted to the FMRI via a weighted regression where weights are chosen with respect to distances to $\psi_j(x)$.

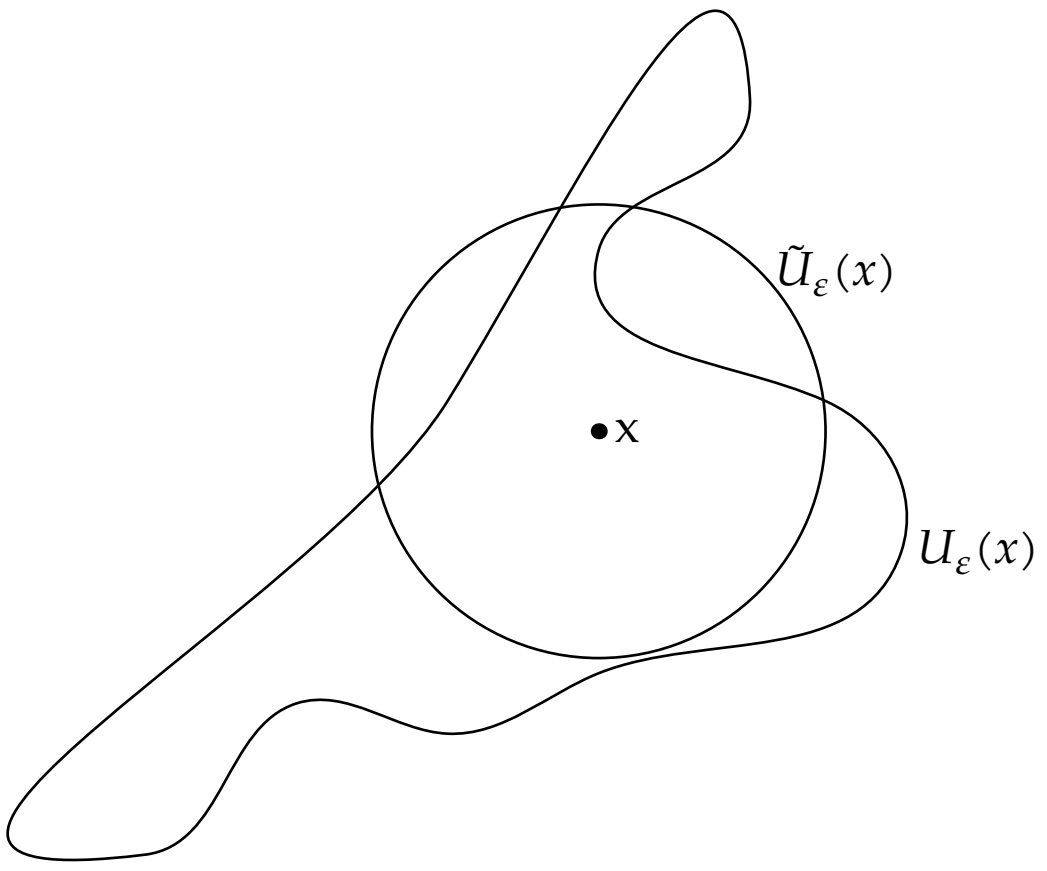

**Figure 4:** Ideally, the same weight is attributed to all points of the boundary of $U_\varepsilon(x)$. This will ordinarily not be the case. If $x$ in this figure is at the centre of a high, task related BOLD effect, then points in $\tilde{U}_\varepsilon(x) \setminus U_\varepsilon(x)$ will decrease the power for detecting this effect at $x$ as the tails of the weighting scheme that extend into $\tilde{U}_\varepsilon(x) \setminus U_\varepsilon(x)$ introduce a negative bias to the estimated BOLD effect at $x$. On the other hand, if there exits a tangible BOLD effect in $\tilde{U}_\varepsilon(x) \setminus U_\varepsilon(x)$ but not at $x$, the introduced bias will be positive and the effect at $x$ gets overestimated. The least bias can be expected, if all $\tilde{U}_\varepsilon(x)$ are well contained in $U_\varepsilon(x)$. The sets $\tilde{U}_\varepsilon(x)$, however, cannot be made arbitrarily small, as the space around $x$ is only sampled at a limited resolution. Power can thus be increased by constructing weighting schemes which produce sets $\tilde{U}_\varepsilon(x)$ which approximate the form and extend of the sets $U_\varepsilon(x)$.

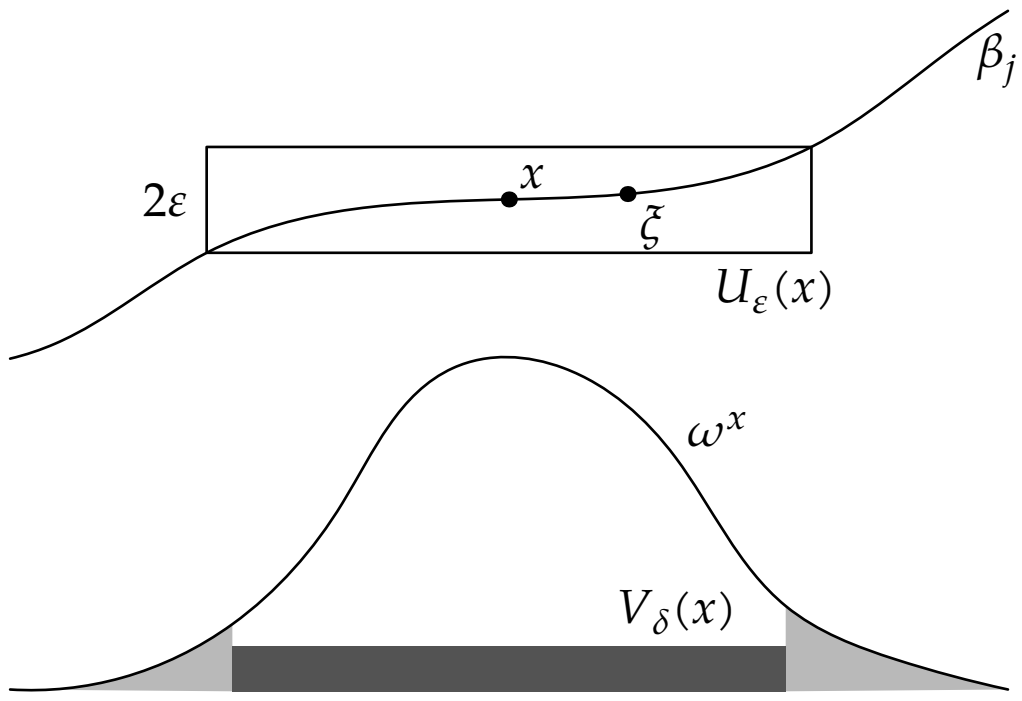

**Figure 5:** A weighting scheme that is $\varepsilon$-nice in $x$. The grey tails of $\omega^x$ are not allowed to be too heavy.

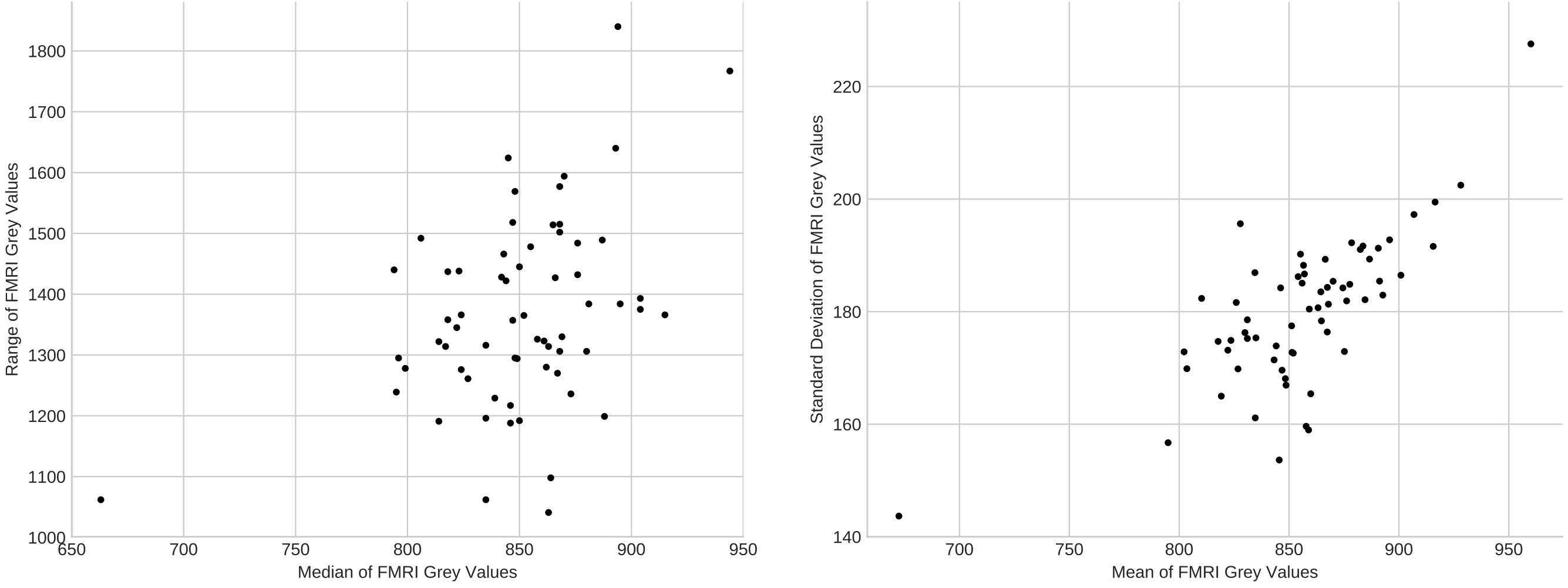

Median and range of FMRI grey values

Mean and standard deviation of FMRI grey values

**Figure 6:** The two plots show the median, range, mean, and standard deviation of grey scale values as they are found in the foreground of a sample of 64 Nifti1 files containing the FMRI data of different subjects. All FMRI have been measured on the same equipment using the same acquisition sequence (see the materials section for details). The plots quite diligently show that the grey values of each FMRI live on their own respective scale. Each scale is transformed such that the grey values fit into the integral data type of the respective format, here the Nifti1 specification. The scales themself have no physical meaning.

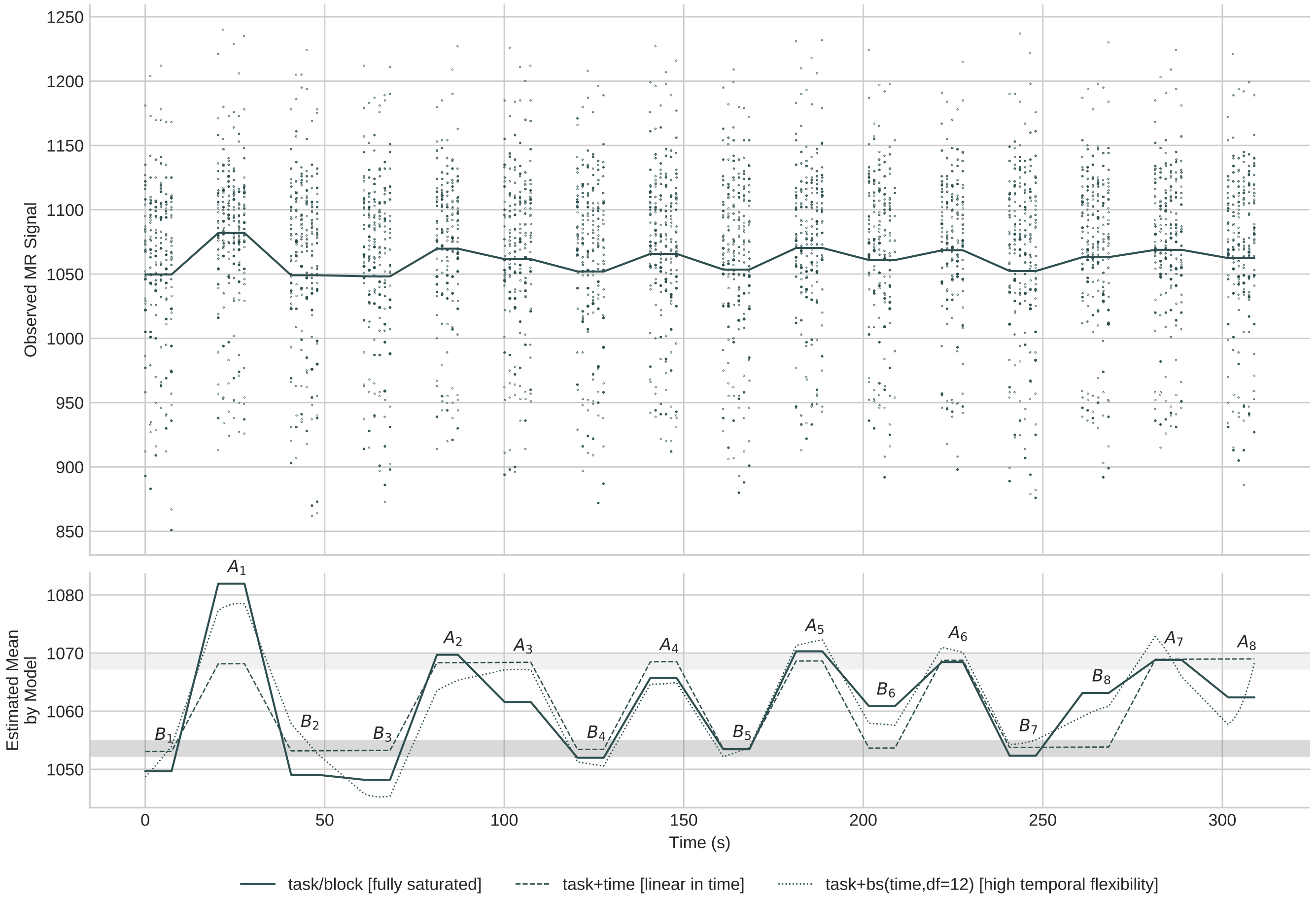

**Figure 7:** All MR measurements of an FMRI session of subject $j_0$ during task blocks within a radius of 6.90 mm around a point $\psi_{j_0}(x_0)$ in the subject are plotted in the top part of the graph. Dots are scaled in size with respect to their distance to $\psi_{j_0}(x_0)$. The smaller, the larger the distance to $\psi_{j_0}(x_0)$. Within block mean intensity is shown as a solid line. The lower part displays the expected mean signal as estimated by three different models. The respective overall block means are viewed as shaded beams.

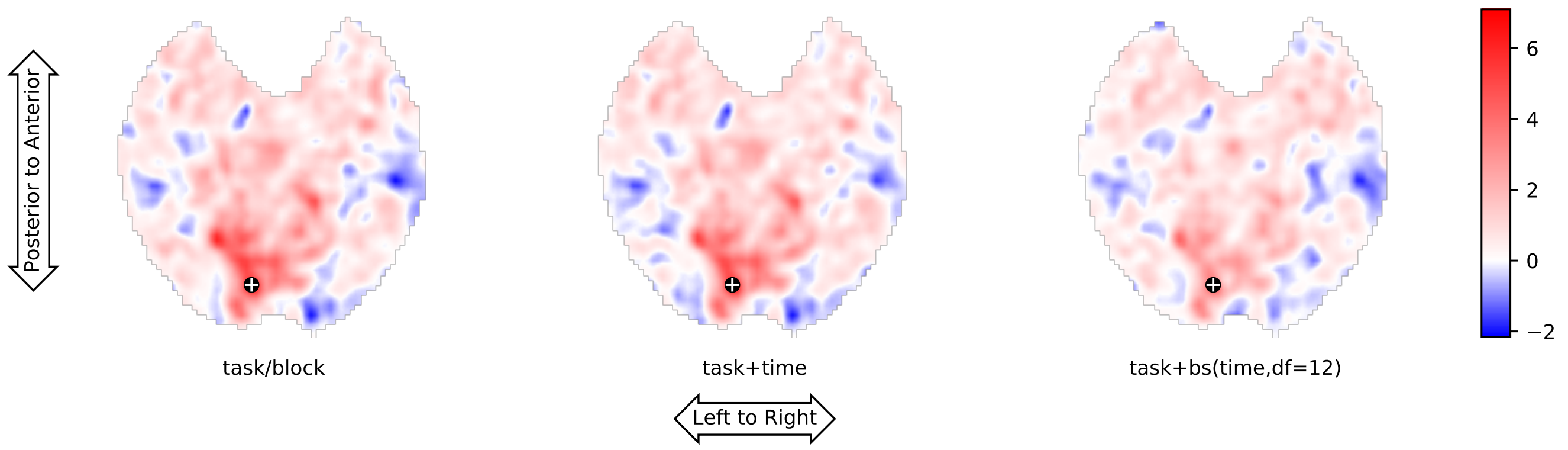

**Figure 8:** Displayed are heat maps of the same transversal cut through three t-statistics fields, each respectively testing for non-zero task effects in the same subject $j_0$ but each with respect to a different model. Fields are shown in $(2\text{ mm})^3$ resolution. The common point $x_0$, which is located in the occipital lobe of the subject and marked by a cross in the maps, shows the highest peak in each of the three 3D fields. This suggests that $x_0$ lies at the centre of an area of high, task specific BOLD activity. The actual measurements around $x_0$ are plotted in Figure 7. A loss in power is visible: from left to right, the images are getting more pale and less replete.

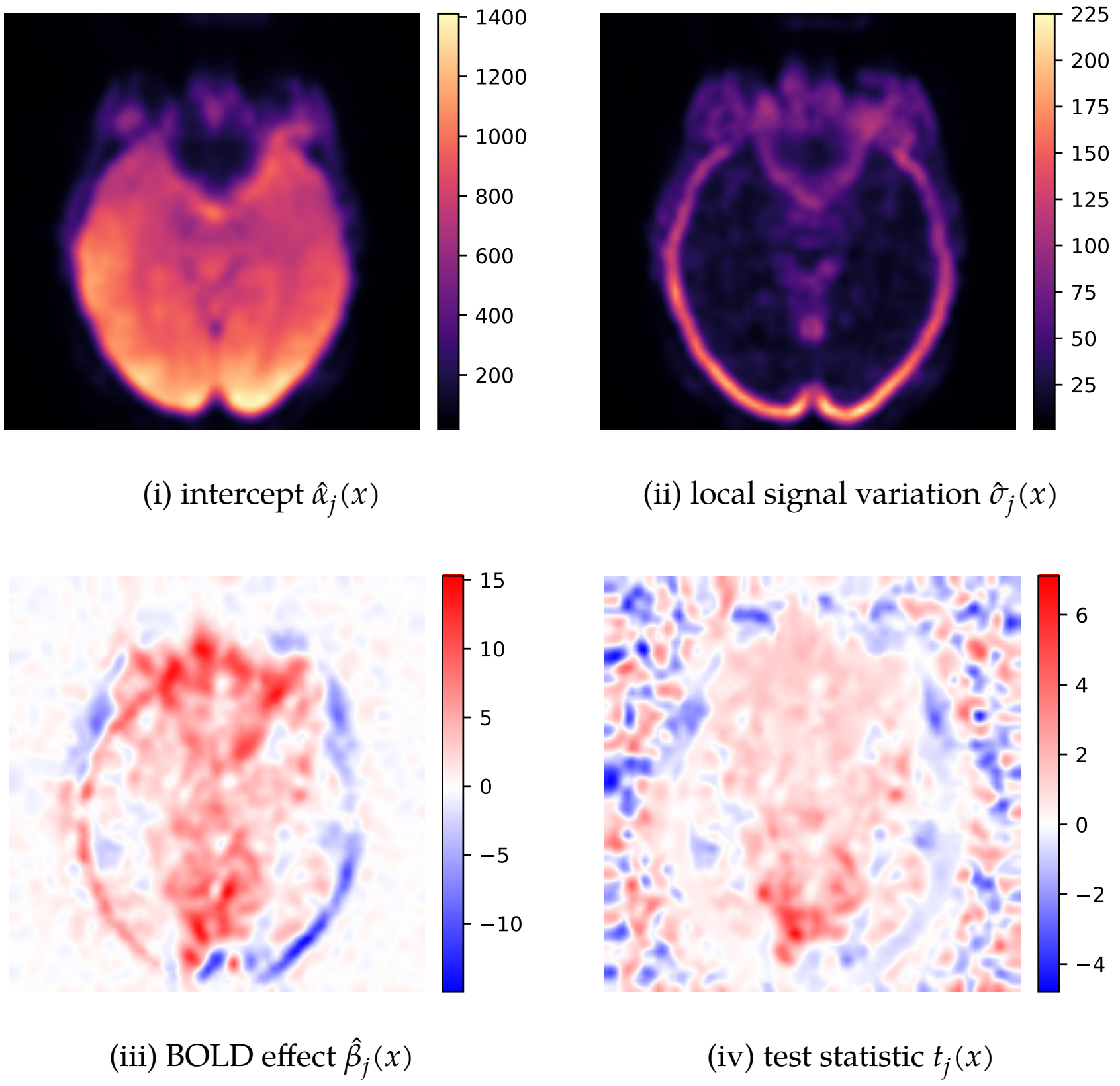

(i) intercept $\hat{\alpha}_j(x)$ (ii) local signal variation $\hat{\sigma}_j(x)$

(iii) BOLD effect $\hat{\beta}_j(x)$ (iv) test statistic $t_j(x)$

**Figure 9:** A saturated, nested model was fitted to the FMRI data of the subject $j_0$ without any prior foreground / background separation of the respective images. Plotted are heat maps of various parameter fields of the model: (i) the intercept field $\hat{\alpha}_j$, (ii) the local signal variation $\hat{\sigma}_j$ (the standard deviations of the respective residual distributions) (iii) the task specific BOLD effect $\hat{\beta}_j$, and (iv) the t-test statistics field $t_j$ (that tests for non-zero task related BOLD changes of the MR signal). Small errors in the estimated head movements of the subject result in biased estimates of BOLD effect. This is quite visible by the corona of alleged activity of $\hat{\beta}_j$ in (iii). Albeit the large estimates of $\beta_j$ on the surface of the subject's brain, these areas do not show any statistical significance in the test statistics field $t_j$ in (iv) – despite their magnitude. The reason for this is the variance inflation at tissue boundaries and surface areas. Errors in $\hat{\rho}_{jt}$ (the estimated location of the subject in the scanner) lead to a further increase of the residual variance. The inflation is visible in (ii).

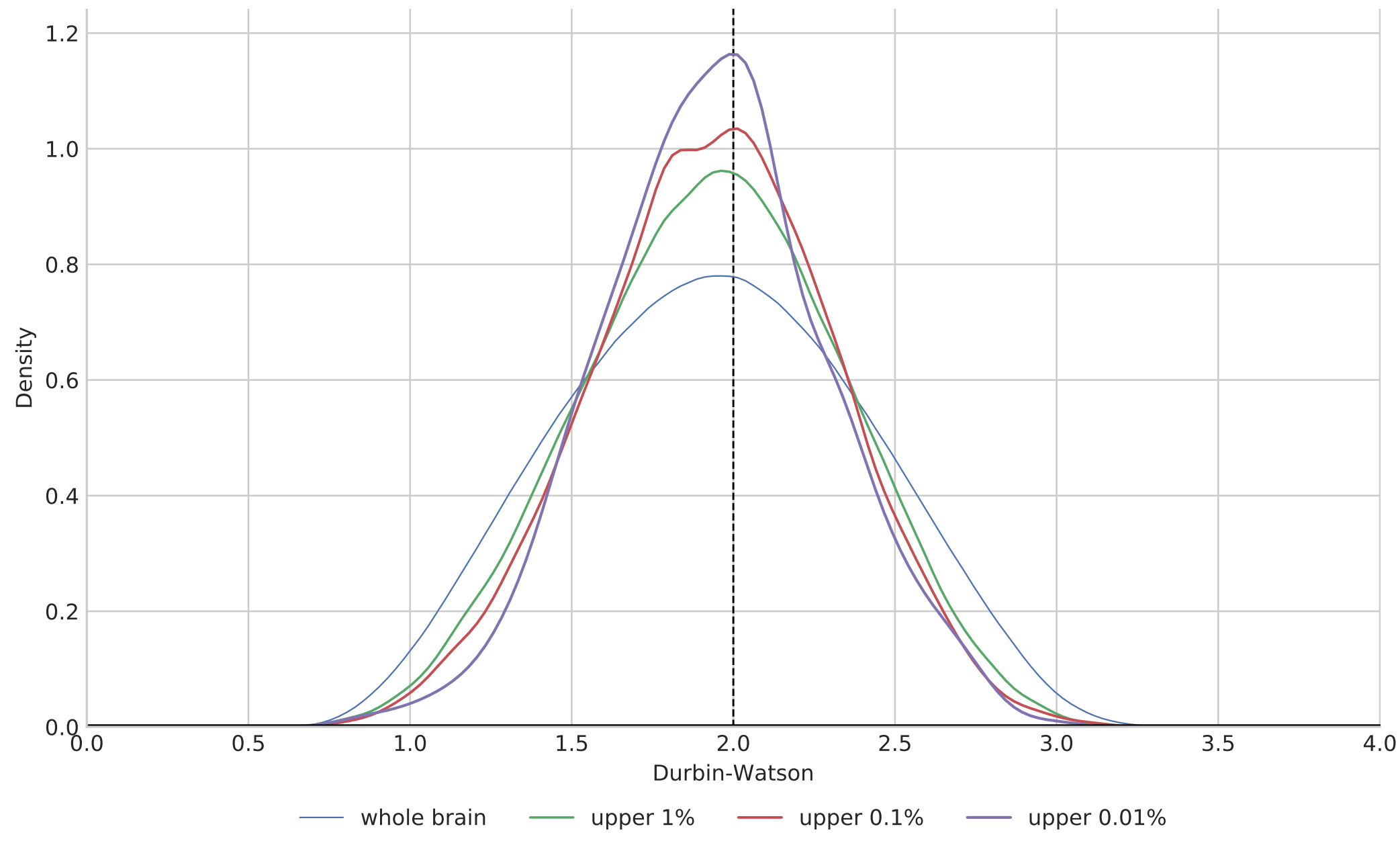

**Figure 10:** Densities of the empirical Durbin-Watson statistics distribution in the brain and in various peak areas of the respective t-statistic field.

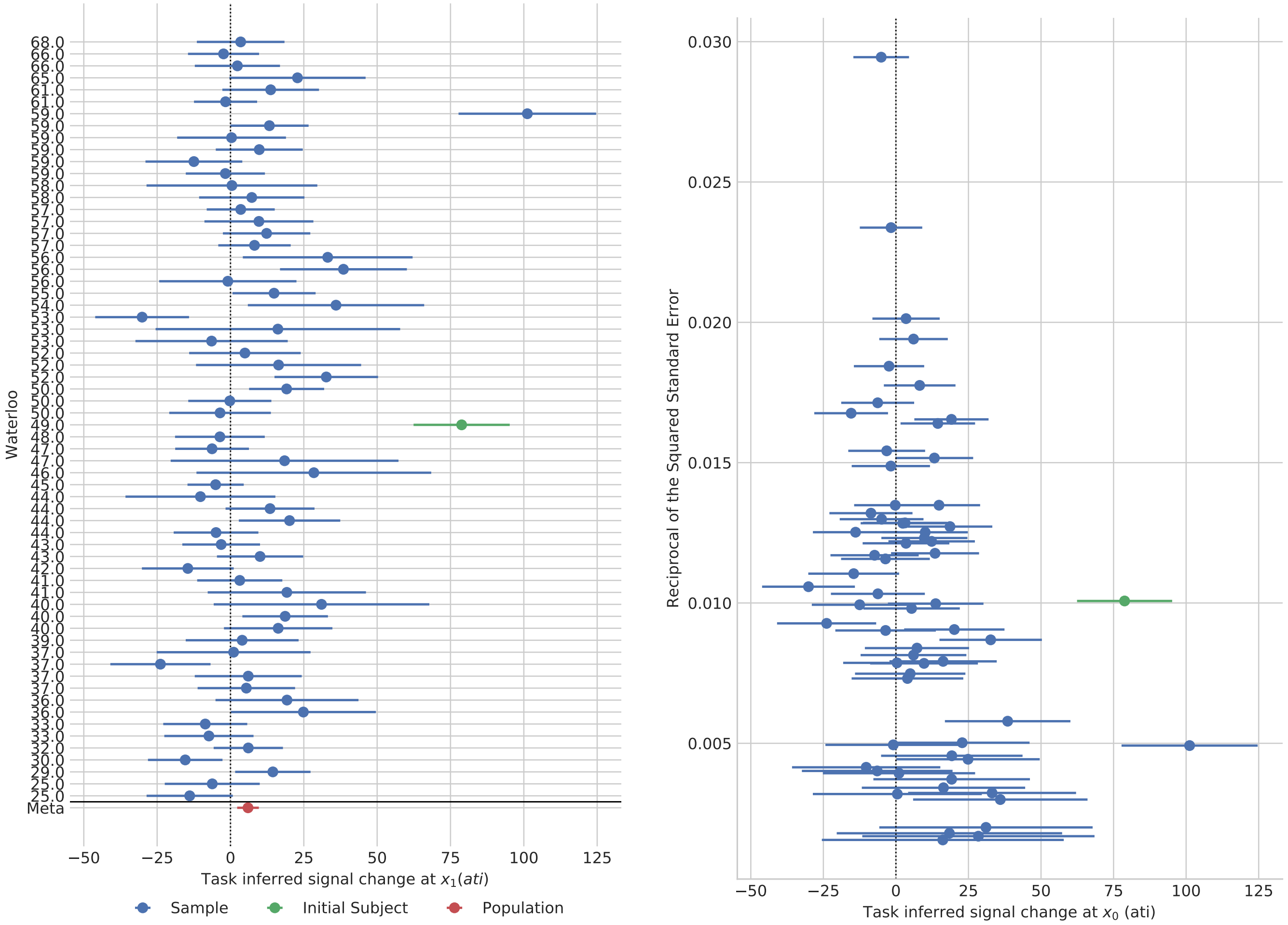

Forest plot at $x_0$ Funnel plot at $x_0$

**Figure 11:** Forest on the left and funnel plot on the right of the estimated BOLD effect at $x_0$ in ati units. The forest plot shows point estimates and 95%-confidence intervals of the respective task related BOLD effect for each subject in the study ordered by handedness on the y-axes. The population inferred BOLD effect is displayed at the bottom. The BOLD effect at $x_0$ in the population is 5.97 ati with $\text{CI}(.95) = [2.29 \text{ ati}, 9.66 \text{ ati}]$ and has a p-value of 0.0044. The population model was fitted using a sample that excluded the initial subject $j_0$ and $x_0$ had been chosen a-priori for this analysis. Neither confidence interval nor p-value for the effect at population level have to be adjusted for multiple testing.

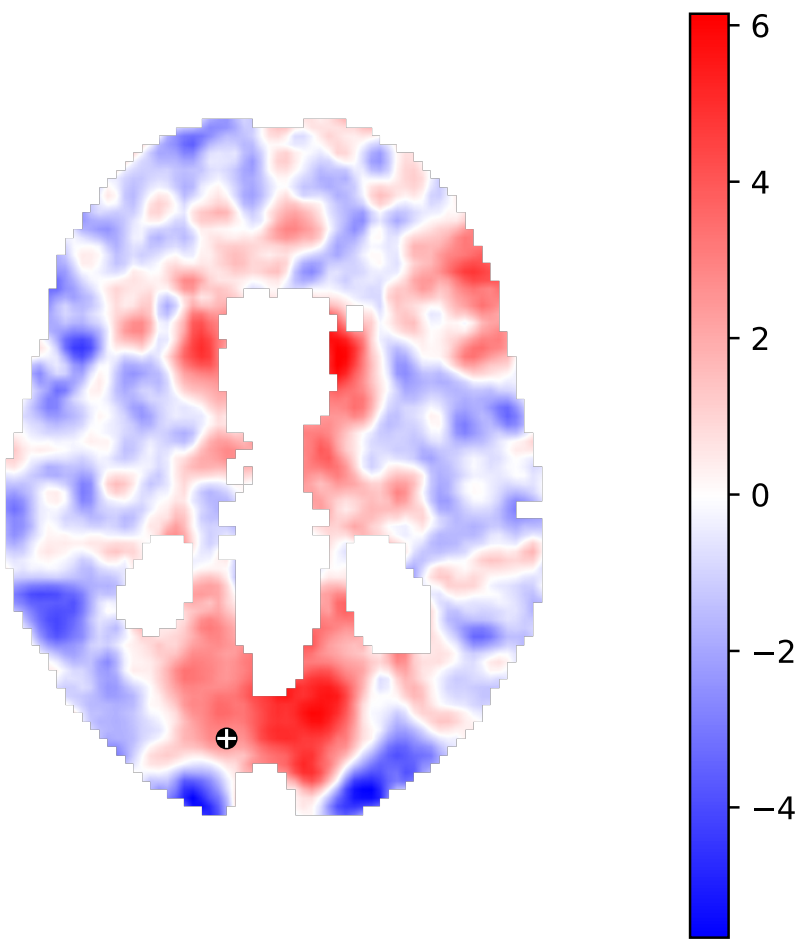

**Figure 12:** Heat map of a transversal cut at $x_0$ through the Knapp-Hartung adjusted t-statistic field that test for non-zero task related BOLD activity in a population of right handed, healthy subjects performing a two-block word generating task. The t-test statistic field is shown in $(2\text{ mm})^3$ resolution in MNI standard coordinates. The fit excluded the initial subject $j_0$. The point $x_0$ lies in the occipital lobe and is marked by a $\oplus$.

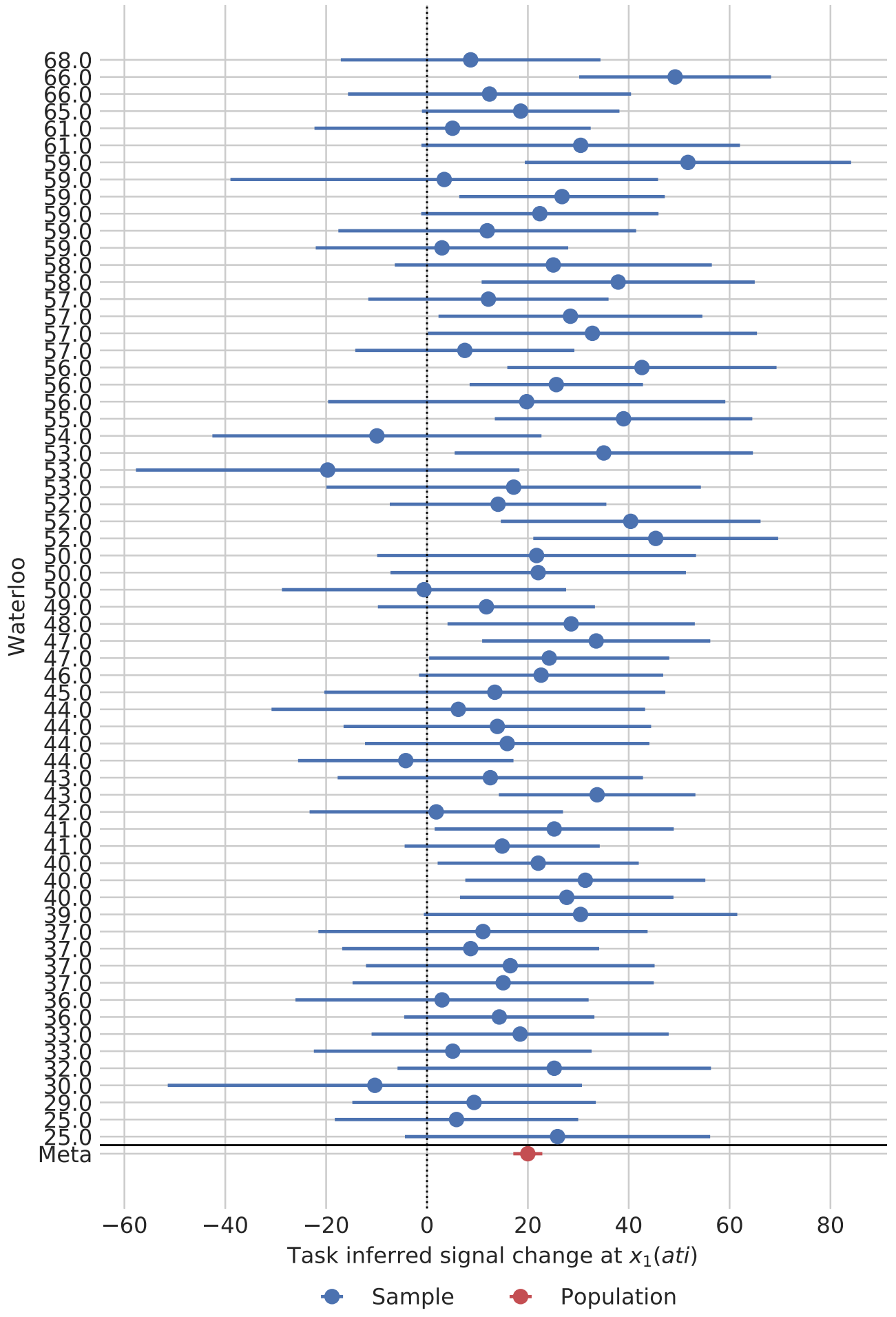

Forest plot at $x_1$

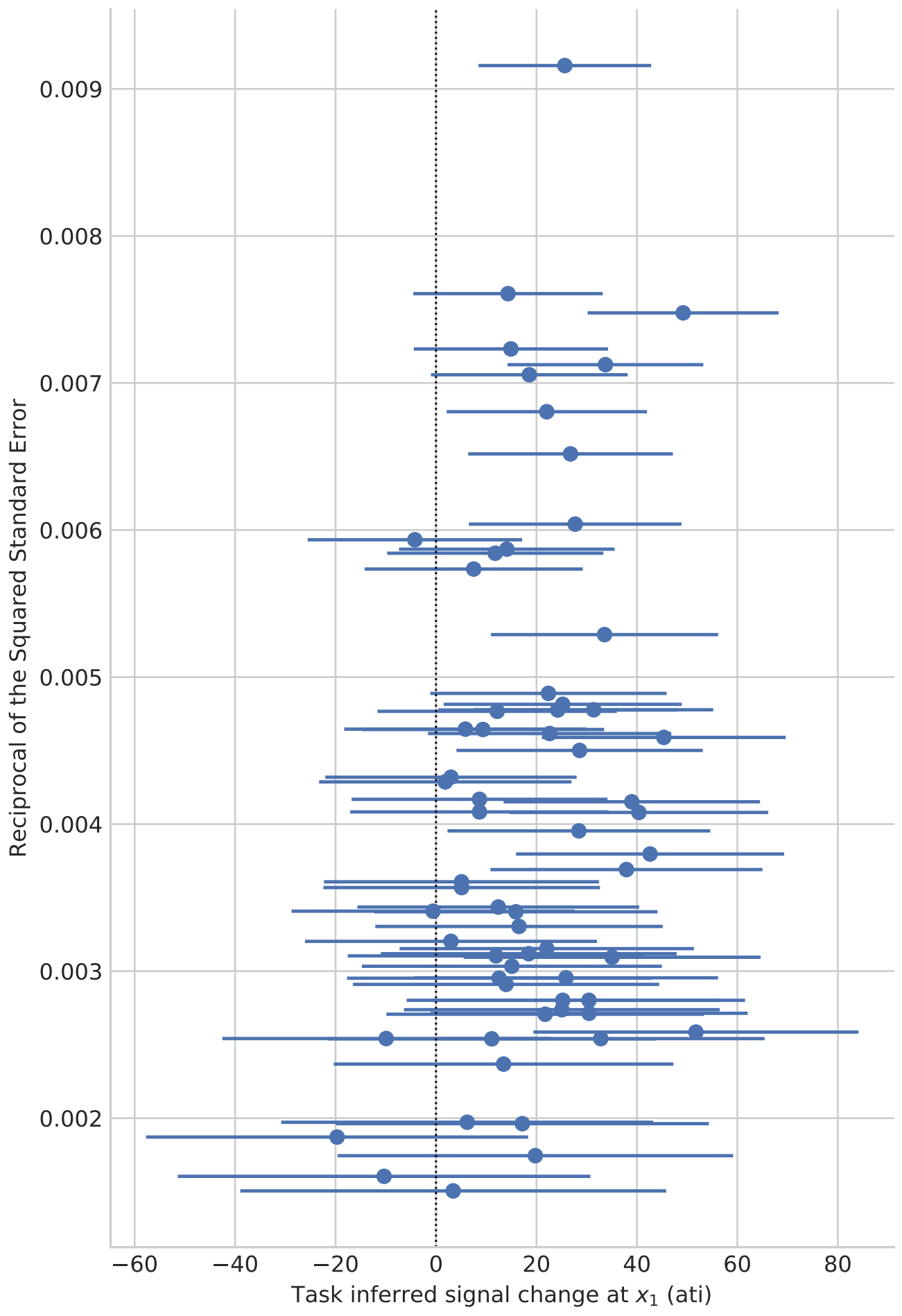

Funnel plot at $x_1$

**Figure 13:** Forest on the left and funnel plot on the right of the estimated BOLD effect at $x_1$ in ati units. The forest plots shows the point estimates together with 95%-confidence intervals of the respective task related BOLD effect at $x_1$ for each subject in the study ordered by handedness on the y-axes. The population inferred BOLD effect is displayed at the bottom of the forest plot. The effect at $x_1$ at population level is 19.99 ati with $\mathrm{CI}(.95) = [17.12 \text{ ati}, 22.87 \text{ ati}]$. The already reduced, Knapp-Hartung adjusted t-statistic is still 11.61 at this point, which corresponds to a p-value well below double-precision floating-points. As the point $x_1$ had been chosen posteriori, the test would need to be adjusted for multiple testing, though.

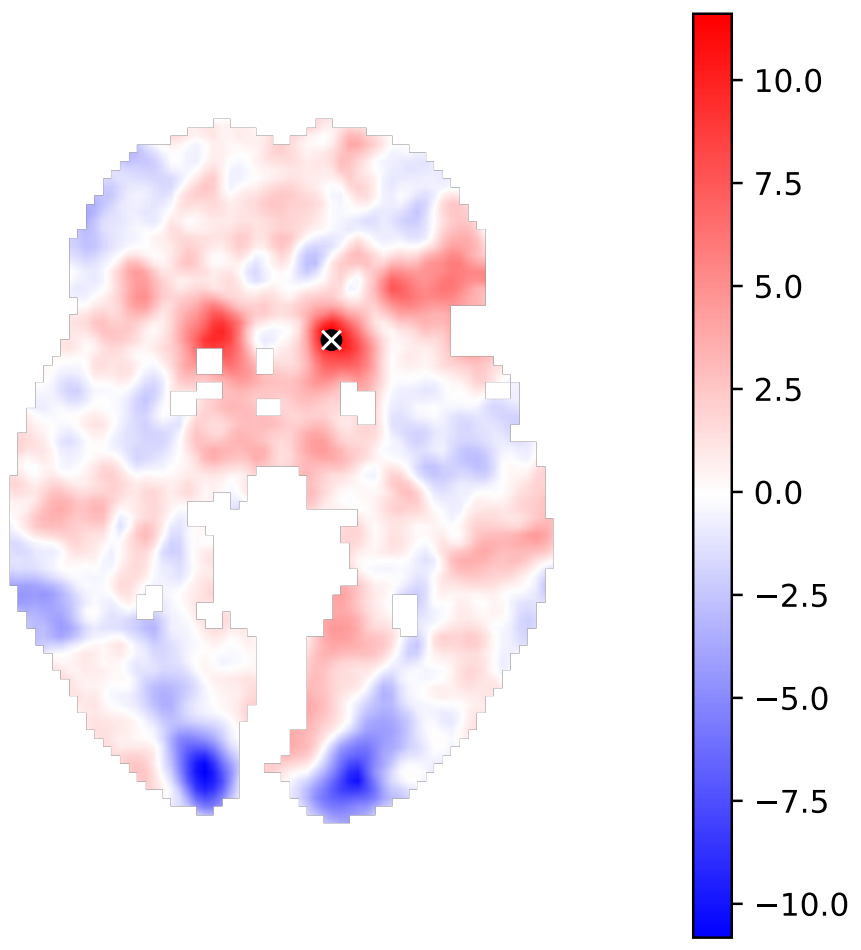

**Figure 14:** Heat map of a transversal cut at $x_1$ through the Knapp-Hartung adjusted t-statistic field that test for non-zero task related BOLD activity in a population of right handed, healthy subjects performing a two-block word generating task. The t-test statistic field is shown in $(2 \text{ mm})^3$ resolution in MNI standard coordinates. The fit included the initial subject $j_0$. The point $x_1$ lies in the caudate and is marked by a ⊗.

1. Poline, J.-B. and Brett, M., The general linear model and fMRI: does love last forever?, *NeuroImage* **62**(2), 439-464 (2012).
2. Lindquist, M.A., The statistical analysis of fMRI data, *Statistical Science* **23**(4), 439-464 (2008).
3. Brown, E. and Behrmann, M., Controversy in statistical analysis of functional magnetic resonance imaging data, *Proceedings of the National Academy of Sciences* **114**(17), E3368-E3369 (2017).
4. Eklund, A., Nichols, T.E., and Knutsson, H., Cluster failure: Why fMRI inferences for spatial extent have inflated false-positive rates, *Proceedings of the National Academy of Sciences* **113**(28), 7900-7905 (2016).
5. Friston, K.J., Ashburner, J., Kiebel, S., Nichols, T., and Penny, W., *Statistical Parametric Mapping: The Analysis of Functional Brain Images* (Elsevier/Academic, London, 2007).
6. Ashburner, J., SPM: A history, *NeuroImage* **62**(2), 791-800 (2012).
7. Jenkinson, M., Beckmann, C.F., Behrens, T.E., Woolrich, M.W., and Smith, S.M., FSL, *NeuroImage* **62**(2), 782-790 (2012).
8. Cox, R.W., AFNI: Software for analysis and visualization of functional magnetic resonance neuroimages, *Computers and Biomedical Research Journal* **29**(3), 162-173 (1996).
9. Polzehl, J. and Spokoiny, V.G., Adaptive weights smoothing with applications to image restoration, *Journal of the Royal Statistical Society* **62**(2), 335-354 (2000).
10. Carp, J., The secret lives of experiments: methods reporting in the fMRI literature, *NeuroImage* **63** 289-300 (2012).
11. Carp, J., On the plurality of (methodological) worlds: estimating the analytic flexibility of fMRI experiments, *Frontiers in Neuroscience* **6** 1-12 (2012).
12. Friston, K.J., Worsley, K.J., Frackowiak, R.S.J., Mazziotta, J.C., and Evans, A.C., Assessing the significance of focal activations using their spatial extent, *Human brain mapping* **1**(3), 210-220 (1994).
13. Worsley, J.K., Estimating the number of peaks in a random field using the Hadwiger characteristic of excursion sets, with applications to medical images, *The Annals of Statistics* **23** 640-669 (1995).
14. Worsley, J.K., Marrett, S., Neelin, P., Vandal, A.C., Friston, K.J., and Evans, A.C., A unified statistical approach for determining significant signals in images of cerebral activation, *Human brain mapping* **4**(1), 58-73 (1996).
15. Berkey, C.S., Hoaglin, D.C., Mosteller, F., and Colditz, C.A., A random-effects regression model for meta-analysis, *Statistics in Medicine* **14** 395-411 (1995).
16. DerSimonian, R. and Laird, N., Meta-analysis in clinical trials, *Controlled Clinical Trials* **7**(3), 177-188 (1986).
17. Hedges, L.V., A random effects model for effect sizes, *Psychological Bulletin* **93**(2), 388-395 (1983).
18. Hedges, L.V. and Olkin, I., *Statistical Methods for Meta-Analysis* (Academic Press, Orlando, 1985).
19. Sidik, K. and Jonkman, J.N., Simple heterogeneity variance estimation for meta-analysis, *Applied Statistical Science* **54** 367-384 (2005).
20. Paule, R.C. and Mandel, J., Consensus values and weighting factors, *Journal of Research of the National Bureau of Standards* **87** 377-385 (1982).
21. Friedrich, T.W.D. and Knapp, G., Generalised Interval Estimation in the Random Effects Meta Regression Model, *Computational Statistics and Data Analysis* **64** 165-179 (2013).
22. Knapp, G. and Hartung, J., Improved tests for a random effects meta-regression with a single covariate, *Statistics in Medicine* **22** 2693-2710 (2003).
23. Kutner, M.H., Nachtsheim, C.J., Neter, J., and Li, W., *Applied Linear Statistical Models* (McGraw-Hill/Irwin, New York, 2005).
24. Woolrich, M., Ripley, B.D., Brady, M., and Smith, S.M., Temporal autocorrelation in univariate linear modeling of FMRI data, *NeuroImage* **14** 1370-1386 (2001).
25. Durbin, J. and Watson, G.S., Testing for Serial Correlation in least Squares Regression: I, *Biometrika* **37**(3), 409-428 (1950).
26. Siegmund, D.O. and Worsley, K.J., Testing for a signal with unknown location and scale in a stationary Gaussian random field, *The Annals of Statistics* **23**(2), 608-639 (1995).
27. Cao, J. and Worsley, K.J., The detection of local shape changes via the geometry of Hotelling's $t^2$ fields, *The Annals of Statistics* **27**(3), 925-942 (1999).
28. Munafo, M.R., Nosek, B.A., Bishop, D.V.M., Button, K.S., Chambers, C.D., du Sert, N.P., Simonsohn, U., Wagenmakers, E.-J., Ware, J.J., and Ioannidis, J.P.A., A manifesto for reproducible science, *Nature Human Behaviour* **1**(21), 1-9 (2017).
29. Brett, M., Hanke, M., Markiewicz, C., Côté, M.-A., McCarthy, P., Ghosh, S., Wassermann, D., Gerhard, S., Halchenko, Y., Larson, E. *et al.*, *Nipy/nibabel: 2.3.0*, (2018).
30. Otsu, N., A Threshold Selection Method from Gray-Level Histograms, *IEEE Transactions on Systems, Man, and Cybernetics* **9**(1), 62-66 (1979).
31. Grubbs, F.E., Sample criteria for testing outlying observations, *Annals of Mathematical Statistics* **21**(1), 27-58 (1950).
32. Avants, B.B., Tustison, N.J., Song, G., Cook, P.A., Klein, A., and Gee, J.C., A reproducible evaluation of ANTs similarity metric performance in brain image registration, *NeuroImage* **54**(3), 2033-2044 (2011).